\documentclass[fleqn,usenatbib,onecolumn]{mnras}

\usepackage{txfonts}
\usepackage{refcount}
\usepackage[T1]{fontenc}
\usepackage{rotating}
\usepackage[percent]{overpic}
\DeclareRobustCommand{\VAN}[3]{#2}
\let\VANthebibliography\thebibliography
\def\thebibliography{\DeclareRobustCommand{\VAN}[3]{##3}\VANthebibliography}

\usepackage{xcolor}

\defcitealias{Cheng89}{CR89} 

\usepackage{graphicx}	
\usepackage{amssymb}	

\title[GeV$-$TeV $\gamma-$rays from accreting neutron stars]{Modelling the expected very high energy $\gamma$-ray emission from accreting neutron stars in X-ray binaries}

\author[L. Ducci et al.]{
L. Ducci,$^{1,2,3}$\thanks{E-mail: ducci@astro.uni-tuebingen.de (LD)}
P. Romano,$^{3}$
S. Vercellone,$^{3}$
A. Santangelo$^{1}$
\\
$^{1}$Institut f\"ur Astronomie und Astrophysik, Kepler Center for Astro and Particle Physics, University of T\"ubingen, 
              Sand 1, 72076 T\"ubingen, Germany\\
$^{2}$ ISDC Data Center for Astrophysics, Universit\'e de Gen\`eve, 16 chemin d'\'Ecogia, 
                  1290 Versoix, Switzerland\\
$^{3}$INAF -- Osservatorio Astronomico di Brera, 
                  Via E.\ Bianchi 46, I-23807, Merate, Italy
}

\date{Accepted XXX. Received YYY; in original form ZZZ}

\pubyear{2023}

\begin{document}
\label{firstpage}
\pagerange{\pageref{firstpage}--\pageref{lastpage}}
\maketitle

\begin{abstract}
The detection of $\gamma-$ray emission from accreting pulsars in X-ray binaries (XRBs) has long been sought after.
For some high-mass X-ray binaries (HMXBs), marginal detections have recently been reported.
Regardless of whether these will be confirmed or not, future telescopes operating
in the $\gamma-$ray band could offer the sensitivity needed to achieve solid detections and possibly spectra.
In view of future observational advances, we explored the expected emission above 10~GeV
from XRBs, based on the Cheng \& Ruderman model, where $\gamma-$ray photons are produced
by the decay of $\pi^0$ originated by protons accelerated in the magnetosphere of an accreting pulsar fed by an accretion disc.
We improved this model by considering, through Monte Carlo simulations, the development of cascades inside of and outside the accretion disc,
taking into account pair and photon production processes that involve interaction with nuclei, X-ray photons from the accretion disc,
and the magnetic field.
We produced grids of solutions for different input parameter values of the X-ray luminosity ($L_{\rm x}$),
magnetic field strength ($B$), and for different properties of the region where acceleration occurs.
We found that the $\gamma-$ray luminosity spans more than five orders of magnitude, with a maximum of $\sim10^{35}$~erg~s$^{-1}$.
The $\gamma-$ray spectra show a large variety of shapes: some have most of the emission below $\sim100$~GeV,
others are harder (emission up to 10$-$100~TeV).
We compared our results with \emph{Fermi}/LAT and VERITAS detections and upper-limits
of two HMXBs: A0535+26 and GRO~J1008$-$57. More consequential comparisons will be possible when
more sensitive instruments will be operational in the coming years.
\end{abstract}




\section{Introduction}
\label{Sect. intro}

X-ray binaries (XRBs) are systems usually composed by
a compact object (neutron star, black hole, white dwarf) and a
normal star (often called ``donor star'').
The X-ray emission is produced by the conversion of the gravitational
potential energy of the accreted matter into radiation \citep[see, e.g., ][]{Shklovsky67}.
XRBs are usually divided in two main classes: high-mass X-ray binaries (HMXBs)
and low-mass X-ray binaries (LMXBs).
In HMXBs, the donor star is massive ($\gtrsim 8$~M$_\odot$) and of OB spectral type.
The compact object accretes the dense stellar wind ejected by the companion star,
and the accretion can be either spherically symmetric or mediated by an accretion disc.
In LMXBs, the donor star is less massive than the Sun and usually it transfers mass
by Roche Lobe overflow. Usually, an accretion disc forms around the compact object
\citep[see, e.g., ][]{Davidson73, Treves88, vanParadijs98}.
For both classes, X-ray emission can be either persistent or transient.
X-ray luminosity ranges from $\sim 10^{31}$~erg~s$^{-1}$ to $\sim 10^{39}$~erg~s$^{-1}$,
and the bulk of the emission is in the band $\sim 0.1-100$~keV \citep[see, e.g., ][]{Revnivtsev15}.
Within this energy range, spectra can be softer or harder, depending on many physical
conditions \citep[e.g.][]{Makishima86, White95}.
The orbital periods range from few hours (for example, the ultra-compact binaries, \citealt{DiSalvo22})
to few years \citep[e.g. ][]{Bhadkamkar12}.

Since the birth of $\gamma$-ray astronomy, XRBs received special attention,
because they were considered candidates for the origin of a substantial fraction of the Galactic cosmic rays.
In particular, it has been proposed that particles accelerated in XRBs might contribute to the Galactic cosmic-ray sea \citep{Heinz02}.
Part of the interest on XRBs was fueled further by the alleged $\sim 40$ detections obtained between 1975 and 1990
at energies $\sim$35~MeV$-5$~GeV (SAS-2) and around 1~TeV in the directions of
Her~X$-$1, Vela~X$-$1, Cen~X$-$3, 4U~1145$-$619, 4U~0115+63, 4U~1626$-$67, Sco~X$-$1, Cyg~X$-$3, SMC~X$-$1, LMC~X$-$4,
and X0021.8$-$7221 \citep[][ and references therein]{Weekes88, Bonnet-Bidaud88, Chadwick90, Ong98}.
Many of these detections were marginally significant and were not confirmed
in the subsequent years by more sensitive experiments,
casting doubts on their truthfulness \citep{Ong98, Dubus13}.

In the late 1970s, the discovery of X-ray emission from the Be/X-ray binary (Be/XRB) LS~I+61$^\circ$303
was also important (although it played a much more fundamental role in the birth of the class of
``gamma-ray binaries'', that finally emerged in 2000s; \citep[see, e.g., ][ and references therein]{Dubus13}.
$\gamma$-ray binaries host a compact object (pulsar or black-hole) and a massive OB type star.
In gamma-ray binaries, the bulk of the emission is in the $\gamma-$ray band.
They also show characteristic variability in the radio band.
The main physical processes proposed for their  $\gamma-$ray emission are particles accelerated
in relativistic jets (microquasar scenario) or interaction between the winds of a pulsar and the companion star
\citep{Dubus06, Mirabel06, Romero07, Paredes19}.
Their relatively weak X-ray emission and their overall multiwavelength properties
suggest they form a distinct class from the typical XRBs, for which most of the emission in the X-ray band
is due to accretion. Therefore, $\gamma-$ray binaries are not the subject of this paper.

Following the controversial XRB findings mentioned above, thanks to the advent of new instruments such as
the Energetic Gamma Ray Experiment Telescope (EGRET, 30~MeV$-$10~GeV) on board of the Compton Gamma Ray Observatory (CGRO) satellite \citep{1988SSRv...49...69K},
the Large Area Telescope (LAT) on board of the \emph{Fermi} satellite ($\sim$20~MeV$-$300~GeV; \citealt{2009ApJ...697.1071A}),
and the Astrorivelatore Gamma a Immagini LEggero \emph{AGILE} (30~MeV$-$50~GeV; \citealt{2009A&A...502..995T}),
recently there have been other detections of XRBs (which do not fall into the category of $\gamma-$ray binaries).
For some of them the association between the X-ray and the $\gamma-$ray sources is circumstantial,
being based only on positional coincidence.
For others, a detection (often only marginal) of the $\gamma-$ray source is reported at some orbital phases or specific time intervals
which correspond to epochs in which the X-ray counterpart is in a particular luminosity state (e.g. during X-ray outbursts, or in their temporal proximity).
Table \ref{tab:list_sources} shows a list of the sources recently detected, which are of interest for our work.
From the list of sources reported in \citet{Harvey22} we did not consider the ``false positive'' sources and those
which have an unlikely association with the X-ray counterpart.

All the $\gamma-$ray detections with potential XRBs as counterparts collected since 1975
stimulated the development of numerous hypotheses and models to explain the $\gamma-$ray emission produced in accreting XRBs.
In order to offer a compact review of all the models that, to our knowledge, have been presented so far,
and to put our calculations (Sect. \ref{Sect. Calculations}) in the broader context of all the models proposed, 
we summarised them below.
Since our calculations are based on \citet{Cheng89} (hereafter: \citetalias{Cheng89}) model, this is described more extensively in
Sect. \ref{sect. intro CR}, while the other models are reported in Sect. \ref{Sect. Models}.

\begin{table*}
\begin{center}
\caption{List of XRBs with a $\gamma$-ray counterpart candidate of interest for this work.}
\label{tab:list_sources}
\begin{tabular}{lccclll} 
\hline
Name               &   Type$^a$    & $P_{\rm spin}$   &         B         & Name $\gamma-$ray counterpart       & Type of association$^b$    & References   \\ 
                   &               &    (s)         &        (G)       &                                      &                            &              \\
\hline
A0535+26           &   Be/XRB      &   103.4       & $4\times 10^{12}$  &   3EG~J0542+2610                     &   p,t,o                    &  \getrefnumber{Romero01}, \getrefnumber{Torres01}, \getrefnumber{Harvey22}, \getrefnumber{Hou23} \\
GRO J1008$-$57     &   Be/XRB      &    93.6       & $7-8\times 10^{12}$&  -                                   &   p,t,o                    &  \getrefnumber{Xing19}, \getrefnumber{Harvey22}\\
4U 1036$-$56       &   Be/XRB      &   853         &  -                & AGL~J1037$-$5708; GRO~J1036$-$55     &   p                        &  \getrefnumber{Li12}\\
IGR J17354$-$3255  &    SFXT       &     -         &  -                & AGL~J1734$-$3310                     &   p                        &  \getrefnumber{Sguera11}, \getrefnumber{Garcia14}\\
IGR J11215$-$5952  &    SFXT       &    186        &  -                & 3EG~J1122$-$5946                     &   p                        &  \getrefnumber{Sguera09b}, \getrefnumber{Garcia14}\\
AX J1841.0$-$0536  &    SFXT       &     -         &   -               & 3EG~J1837$-$0423, HESS~J1841$-$055   &   p                        &  \getrefnumber{Sguera09},  \getrefnumber{Garcia14}\\
SAX~J1324.4$-$6200 &  Be/XRB       &   172.8       &   -               & -                                    &   p                        & \getrefnumber{Harvey22}\\
1H~0749$-$600      &  Be/XRB       &     -         &   -               &  -                                   &   p                        & \getrefnumber{Harvey22}\\
1H~1238$-$599      &   HMXB        &   3.2 or 4.3  &   -                &  -                                  &   p                        & \getrefnumber{Harvey22}\\ 
IGR J17544$-$2619  &   SFXT        &    -          & $1.5\times 10^{12}$ & -                                   &   p                        & \getrefnumber{Harvey22}\\
RX J2030.5+4751    &   Be/XRB      &    -          &  -                &  -                                   &   p                        & \getrefnumber{Harvey22}\\
\hline
\end{tabular}
\end{center}
\begin{flushleft}
  $^a$ Sub-classes of XRBs. Be/XRB indicates the subclass of HMXBs with a Be type donor star \citep[see, e.g., ][]{Reig11}. SFXT is for: supergiant fast X-ray transient, a subclass of HMXBs \citep[see, e.g., ][]{Romano15,Sidoli13};
  $^b$ positional: $p$; time variability correlation: $t$; orbital variability correlation: $o$.\\
  References: 
\newcounter{ctr_tabrefs} 
\setrefcountdefault{-99}
\refstepcounter{ctr_tabrefs}\label{Romero01}(\getrefnumber{Romero01}) \citet{Romero01};
\refstepcounter{ctr_tabrefs}\label{Torres01}(\getrefnumber{Torres01}) \citet{Torres01};
\refstepcounter{ctr_tabrefs}\label{Harvey22}(\getrefnumber{Harvey22}) \citet{Harvey22};
\refstepcounter{ctr_tabrefs}\label{Hou23}(\getrefnumber{Hou23}) \citet{Hou23};
\refstepcounter{ctr_tabrefs}\label{Xing19}(\getrefnumber{Xing19}) \citet{Xing19};
\refstepcounter{ctr_tabrefs}\label{Li12}(\getrefnumber{Li12}) \citet{Li12};
\refstepcounter{ctr_tabrefs}\label{Sguera11}(\getrefnumber{Sguera11}) \citet{Sguera11};
\refstepcounter{ctr_tabrefs}\label{Garcia14}(\getrefnumber{Garcia14}) \citet{Garcia14};
\refstepcounter{ctr_tabrefs}\label{Sguera09b}(\getrefnumber{Sguera09b}) \citet{Sguera09b};
\refstepcounter{ctr_tabrefs}\label{Sguera09}(\getrefnumber{Sguera09}) \citet{Sguera09}.
\end{flushleft}
\end{table*}

\subsection{Accelerating gaps in magnetospheres of pulsars with accretion discs and production of $\gamma-$ray emission}
\label{sect. intro CR}

To account for the numerous detections of $\gamma-$ray emission from accreting X-ray pulsars,
\citetalias{Cheng89} proposed a mechanism to accelerate particles in the magnetosphere of a pulsar
surrounded by an accretion disc. This mechanism was able to achieve photon energies up to $\sim 100$~TeV.
The rotational and magnetic dipole field axes were assumed to be aligned, and the accretion disc to lie on the equatorial plane of the pulsar.
\citetalias{Cheng89} showed that when the inner part of the accretion disc rotates faster than the pulsar,
a part of the magnetosphere corotates with the pulsar and another part with the sector of the accretion disc to which
it is linked by the magnetic field. These two regions which rotate with different velocities are separated by
gaps which are in principle empty of plasma. A potential drop forms in the gaps and it can reach $\sim 10^{15}-10^{16}$~V.
The gap potential drop can accelerate protons toward the accretion disc, which becomes a target for $\pi^0$ production
and consequently high-energy photons.
\citet{Cheng91b} showed that the $\gamma-$ray photons produced by the $\pi^0$ decay, called ``primary'',
can be converted to $e^\pm$ pairs by the magnetic field, which in turn, produce synchrotron secondary $\gamma-$ray photons.
Assuming the typical parameters for accreting X-ray pulsars, they found that the total $\gamma-$ray power emitted was
just two orders of magnitude lower than the total X-ray power from accretion.
Their results were in good agreeement with the alleged detections obtained in the years 1975-1990 (see Sect. \ref{Sect. intro}).
\citetalias{Cheng89} explained also the short bursts ($\sim 10^3$~s) of $\gamma-$ray emission 
which were claimed to have been observed in some XRBs. It was proposed that they
were the result of $\gamma-$ray photons able to escape from
radially moving annuli in the accretion disc having relatively low local density.
Elsewhere, the accretion disc was sufficiently thick to absorb all the $\gamma-$ray photons produced in the disc.
\citet{Cheng91} extended the study of the properties of the magnetosphere and the accelerating gap
to the cases in which the inner part of the accretion disc has the same or lower rotational velocity of the pulsar.
The model by \citetalias{Cheng89} was based on the approximation that X-ray photons
produced by the accretion at the stellar surface (including the accretion column) are strongly shielded by the accreting matter.
This approximation was supported by the sufficiently high grammage of the material channeled along the
magnetic field lines by the gravitational field, from the inner regions of the accretion disc to the stellar surface.
However, as pointed out by \citet{Cheng91} and \citet{Cheng92}, the shielding effect may not be efficient enough.
In this case (called ``weak shielding'' or ``pair prodution limited''),
some of the X-ray photons from the stellar surface are boosted up to $\gamma-$rays by inverse Compton scattering
with the $e^\pm$ accelerated in the gap. These $\gamma-$rays then transform to $e^\pm$ pairs by collisions with other X-ray photons.
The newly created electrons and positrons are then accelerated by the gap and then boost up other X-ray photons
to higher energies. This mechanism leads to an overall reduction of the potential drop across the gap, whose formulation
was presented in \citet{Cheng91} and \citet{Cheng92}.

The acceleration mechanism originally proposed by \citetalias{Cheng89} included only the hadronic part,
that is the acceleration of protons.
\citet{Cheng92} showed that the proton current directed toward the disc must be balanced by an opposite flow of electrons
that cancels any net charge density from the current flow, and keep steady the potential drop across the gap.
These electrons lose their energies through the development of cascades based on synchrotron, curvature, and photon-photon interactions.
The mechanism continues until the energies of the $\gamma-$ray photons allow the production of pairs of $e^\pm$.
\citet{Cheng92} considered the ``leptonic branch'' in the framework of strongly shielded gap.
In their calculations, the expected $\gamma-$ray emission is steady and softer ($\lesssim 10^{7-8}$~eV) compared to that produced by the
hadronic branch of the \citetalias{Cheng89} model.

\citet{Bednarek93, Bednarek97, Bednarek00} and \citet{Sierpowska05} discussed the role played by the anisotropic thermal radiation fields
emitted by the accretion disc around the pulsar and by the companion star.
They assumed that the spectra of the primary $\gamma-$ray photons or leptons (of unknown origin) were
injected in the binary system, from the compact object, and were propagating in an anisotropic radiation field.
Their calculations considered different angles of observations.
They demonstrated that the observed $\gamma-$ray emission
can show an orbital modulation, and in some cases it can be quenched by the radiation field emitted by the accretion disc.

In a series of papers, \citet{Romero01, Orellana05}, and \citet{Orellana07}, improved the hadronic model first proposed by \citetalias{Cheng89},
always retaining the hypothesis of a strongly shielded gap. Then, they applied it to the case of the Be/XRB A0535+26.
The main innovations introduced were the production of electromagnetic cascades within the accretion disc under the ``approximation A''
(calculations described in \citet{Rossi41}, in which bremsstrahlung and pair prodution from the interactions
of photons with atoms/nuclei are considered, while ionization losses, Compton collisions are neglected and
the cross section for bremsstrahlung and pair production are considered in their relativistic limits).
For the production of cascades outside the accretion disc, they considered pair production 
in photon-photon collisions and Inverse Compton scattering. 
They presented the expected general properties of the $\gamma-$ray spectrum for A0535+26 for some specific cases.
They assumed a pitch angle for the particles
and photons involved in the cascades of 45$^\circ$ with respect to a normal line to the accretion disc plane,
to take into account the angle of misalignment between the spin and magnetic axes, and a beaming factor of 0.3.

\citet{Zhang14} performed numerical calculations of the cascade processes in accreting XRBs in the framework of the leptonic
scenario proposed by \citet{Cheng92} and, as an example, compared their results with the observations of the $\gamma-$ray binary LS~I~$+61^\circ$~303.

In our work, we consider the hadronic model by \citetalias{Cheng89}
in the cases of strongly and weakly shielded gaps, considering more types of interaction processes for the production
of cascades.
In our calculations, we also take into account the role played by the
poloidal magnetic field inside the accretion disc.
Some of the details about the \citetalias{Cheng89} model on which our calculations are strongly linked
are thus described separately from this introductory section, and can be found in Sect. \ref{Sect. Calculations}.

\subsection{Other models for the $\gamma-$ray emission from XRBs}
\label{Sect. Models}

\begin{figure}
\begin{center}
\includegraphics[width=\columnwidth]{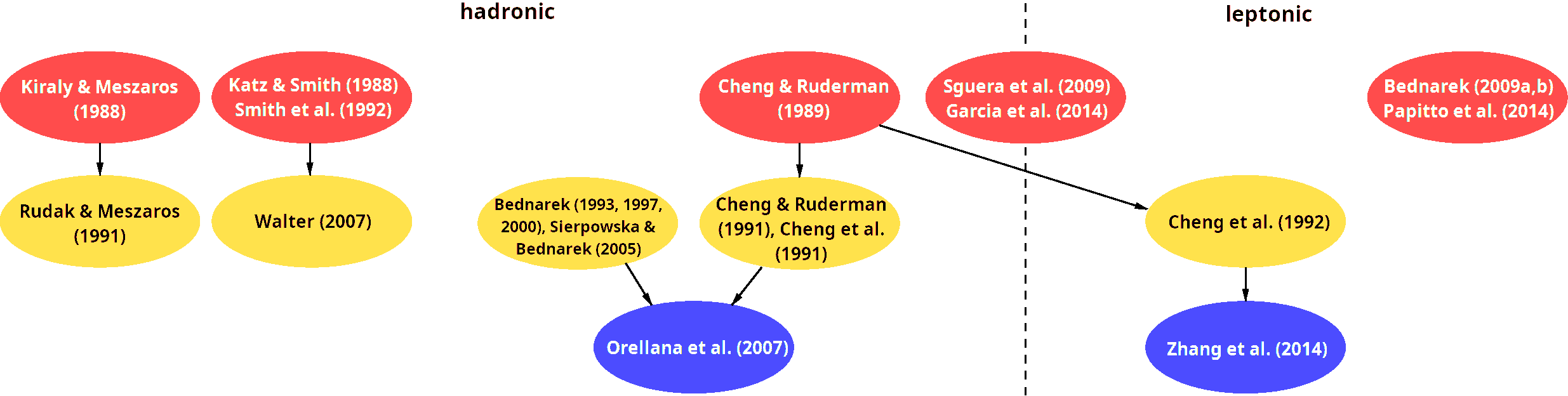}
\end{center}
\caption{Schematic illustration of all the models proposed for the $\gamma$-ray emission from accreting X-ray binaries (to the best of our knowledge).
  They are divided into hadronic and leptonic models. Models in red ellipses include the mechanism of production of particle acceleration and production of primary $\gamma-$ray photons. 
  Models in yellow ellipses assume a mechanism of acceleration of particles and production of primary photons. If they are strictly based on one of them, they are connected with an arrow. Models in blue ellipses are furter development of previous models.}
\label{fig:plot_models}
\end{figure}

In addition to the \citetalias{Cheng89} model and its sub-variants, other ideas and models have been presented so far
for the $\gamma-$ray emission from XRBs. 
Below, we briefly describe them, and in Fig. \ref{fig:plot_models} we show a schematic overview
of all these models, including \citetalias{Cheng89}.

\citet{Katz88} and \citet{Smith92} argued that protons trapped in the
closed magnetosphere of a NS or a white dwarf, bouncing along the dipole magnetic field
lines between the poles, can be accelerated to $10^{16-17}$~eV by plasma turbulences. 
The trapped protons gradually accelerate to high energies, until they  escape the closed magnetosphere. Significant $\gamma-$ray emission occurs
if the protons hit dense regions of matter, such as an accretion disc, the accretion column, or the surface of the companion star.
This mechanism can in principle explain $\gamma-$ray emission up to $\sim 10^{15}$~eV.
\citet{Walter07} showed that if the targets of protons are sufficiently dense clumps of stellar winds expelled by an OB supergiant star companion and accreted by the
compact object, supergiant HMXBs could be transient TeV sources (see, however, also \citealt{Sguera09}).

\citet{Rudak91} calculated the photon spectrum produced by the injection of a power law proton energy distribution
in the magnetosphere of an accreting NS. The photon spectrum is produced by the interaction of protons with the accreting NS X-ray
radiation field and the magnetospheric field.
The results presented in \citet{Rudak91} are dependent on the input proton spectrum, and therefore on the mechanism that accelerates them.
As an example, they carried out the calculations in the framework of the \citet{Kiraly88} corotating jet model, and found that if the input proton energy distribution
is a power law with slope $\Gamma_{\rm p}=2$ and extends up to $E_{\rm p}\approx 10^{16}$~eV, the secondary photon spectrum extends from MeV to the PeV range.

The model described in \citet{Bednarek09b, Bednarek09a} shows that at the Alfv\'en radius a turbulent and magnetized transition region forms, which provides
favorable conditions for the acceleration of electrons to relativistic energies. If a fraction of matter falls onto the neutron star (NS) surface,
it produces thermal radiation which interacts with the accelerated electrons. If accretion is prevented by a gating mechanism,
the relativistic electrons can still interact with the radiation field of the companion star.
Electrons will lose energy through synchrotron and Inverse Compton processes and can produce $\gamma-$ray emission up to the GeV energy range.
This mechanism has been used to discuss the $\gamma-$ray emission from the Be/XRB 4U~1036$-$56, which
has an AGILE and a CGRO counterpart, AGL~J1037$-$5708 and GRO~J1036$-$55 \citep{Li12}.
A similar model was proposed by \citet{Papitto14} to explain the $\gamma-$ray emission of the LMXB XSS J12270-4859.

Positional associations between some supergiant fast X-ray transients
(SFXTs)\footnote{SFXTs are a sub-class of HMXBs where the donor star is OB type supergiant, and the compact object shows a peculiar fast X-ray variability, with a dynamic range up to $10^{5-6}$ on timescales of $\sim 10^4$~s \citep[see, e.g., ][]{Romano15,Sidoli13}.} with unidentified $\gamma-$ray transient sources detected by \emph{AGILE} and EGRET \citep{Sguera09, Sguera09b, Sguera11} have been reported.
\citet{Garcia14} showed that the magnetic field strength in NSs accreting from the stellar wind of a companion OB supergiant star
can decay to sufficiently low values ($B\approx 10^8$~G) to enable the formation of relativistic jets.
For one of them, AX~J1841.0$-$0536, it was proposed that it hosts a low magnetized pulsar where
$\gamma-$ray emission is due to transient jets produced by the sporadic accretion of clumps from the supergiant companion star \citep{Sguera09}.

\section{Calculations}
\label{Sect. Calculations}

\subsection{Mechanism of acceleration of the protons}
\label{Sect. Engine}

\begin{figure}
\begin{center}
  \includegraphics[width=12cm]{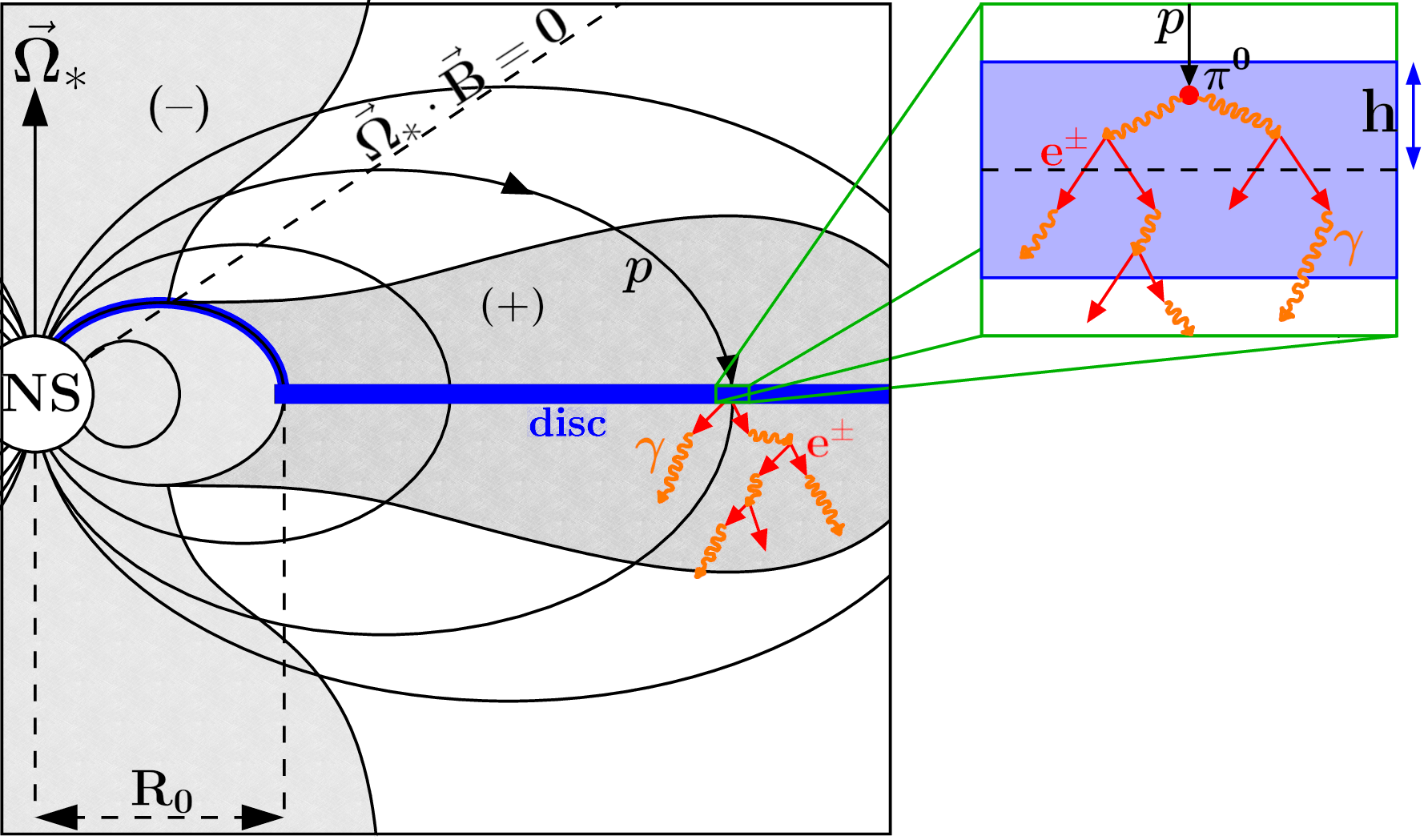}
\end{center}
\caption{Schematic illustration of the Cheng \& Ruderman magnetosphere model (not in scale) when $\Omega_*(R_0)< \Omega_{\rm d}(R_0)$.
  The light gray regions corotate with the NS, while the dark gray regions corotate with the accretion disc.
  A cone-like gap around the null surface $\vec{\Omega}_*\cdot\vec{B}=0$ forms, where protons from the NS are accelerated
  toward the accretion disc. Their interaction with it produce $\gamma-$ray primary photons via $\pi^0$ decay.
  In turn, primary photons develop cascades of $e^\pm$ pairs (red arrows) and other $\gamma-$ray photons (orange wavy arrows)
  inside the accretion disc and in the opposite side of it.}
\label{fig:chengmodel}
\end{figure}

In the \citetalias{Cheng89} model, 
a magnetized NS fed by an accretion disc is considered.
They show that a conelike gap around the null-surface
$\vec{\Omega}_*\cdot\vec{B}=0$ forms when the angular velocity of the disc $\Omega_{\rm d}$
at the inner radius of the disc ($R_0$) exceeds the angular velocity
of the star ($\Omega_*(R_0)< \Omega_{\rm d}(R_0)$).
This gap separates regions with opposite charges, and a strong potential drop develops within it (Fig. \ref{fig:chengmodel}).

If the gap is efficiently shielded by the accretion flow channeled by the magnetic field lines from $\sim R_0$
to the NS surface, it can be considered, in first approximation, empty of plasma.
In this case, called ``strong shielding'', \citetalias{Cheng89} showed that the potential drop across the gap approaches the value:
\begin{equation} \label{eq:DV_strong}
     \Delta V_{\rm strong} \approx 4\times 10^{14} \beta^{-5/2} \left( \frac{M}{M_\odot}\right)^{1/7} R_6^{-4/7} L_{\rm x,37}^{5/7} B_{12}^{-3/7} \mbox{ V\ ,}
\end{equation}
where $M$  is the mass of the NS, $R_6$ (in units of $10^6$\,cm) is its radius,
$B_{12}$ (in units of $10^{12}$\,G) is the magnetic field, $L_{\rm x,37}$ (in units of $10^{37}$\,erg\,s$^{-1}$)
is the X-ray luminosity produced by the accretion, $\beta=2R_0/R_{\rm A}$, where:
\begin{eqnarray}
  R_0      & = &  1.35\gamma_0^{2/7}\eta^{4/7}R_{\rm A}             \label{eq:R0}\\
  R_{\rm A} & = & \mu^{4/7} \dot{M}^{-2/7} (2 G M)^{-1/7} \mbox{\ ,} \label{eq:RA}
\end{eqnarray}
where $R_0$ is defined according to the work by \citet{Wang96}, $R_{\rm A}$ is the Alfven radius,
$\gamma_0=-B_{\phi_0}/B_{z_0}\approx 1$ is the magnetic pitch angle,
$\eta$ is the screening factor (it is assumed a partially screened disc $\eta=0.1$)\footnote{$\eta$ takes into account the screening effects of the currents induced in the surface of the accretion disc.},
$\mu=BR_{\rm NS}^3/2$ is the dipolar magnetic moment, and $\dot{M}$ is the mass accretion rate.

If the gap is weakly shielded,
the potential drop across the gap is:
\begin{equation} \label{eq:DV_weak}
  \Delta V_{\rm weak} \approx 6\times 10^{11} L_{\rm x,37} \left( \frac{E_{\rm x}}{\mbox{1\,keV}}\right)^{-1} R_6^{-1} B_{12}^{-1/2} \mbox{ V\ ,}
\end{equation}
where $E_{\rm x}$ is the characteristic energy of the X-ray photons \citep{Cheng91, Cheng92}.

\begin{figure*}
\begin{center}
  \includegraphics[width=8.5cm]{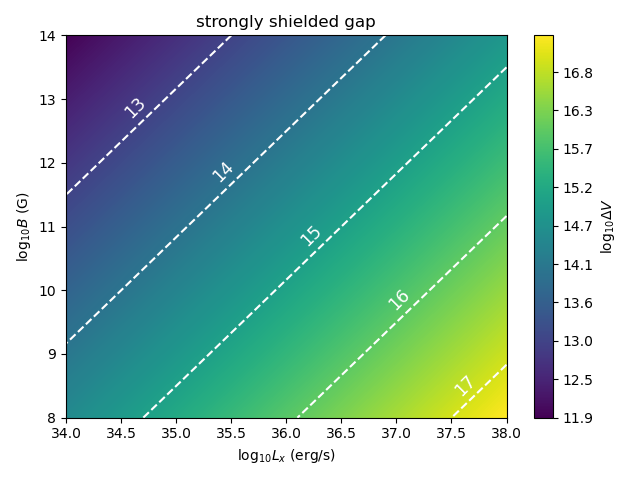}
  \includegraphics[width=8.5cm]{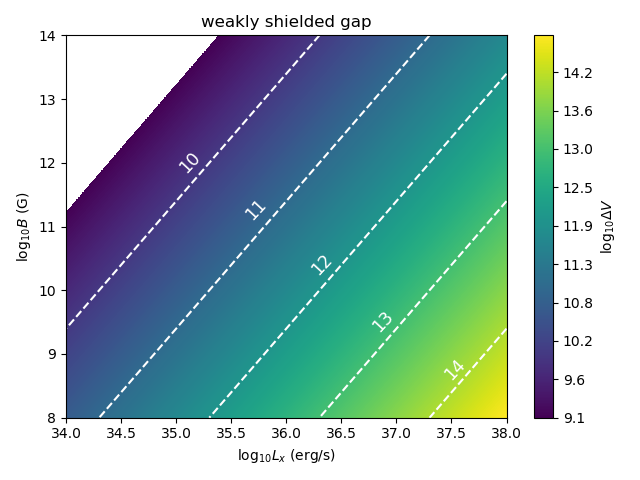}
\end{center}
\caption{Comparison of the potential drops in the gaps for strongly (left) and weakly shielding (right) cases, as function of the X-ray luminosity and magnetic field strength (at the surface of the NS). The white region in the right panel corresponds to $E_{\rm p} < E_{\rm th}$, where the approximations used in this work are not valid Dashed white lines show contours which have the same units used for the respective colour bars.}
\label{fig:deltaV}
\end{figure*}

Figure \ref{fig:deltaV} shows $\Delta V_{\rm strong}$ and $\Delta V_{\rm weak}$ for different values of $L_{\rm x,37}$ and $B_{12}$
(Equations \ref{eq:DV_strong} and \ref{eq:DV_weak}). We assumed $M/M_\odot=1.4$ and $R_6=1.2$.

\begin{figure}
  \begin{center}
    \includegraphics[width=8.5cm]{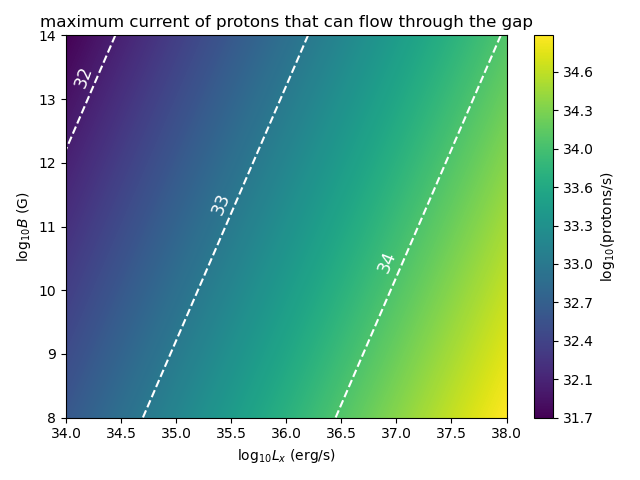}    
\end{center}
\caption{Maximum current of protons that can flow through the gap, as function of the X-ray luminosity and magnetic
field strength (at the surface of the NS). Dashed white lines show contours which have the same units used for the colour bar.}
\label{fig:Jmax}
\end{figure}

The maximum current of protons that can flow through the gap is \citepalias{Cheng89}:
\begin{equation} \label{eq:Jmax}
  J_{\rm max} \approx 1.5\times 10^{24} \beta^{-2} \left(\frac{M}{M_\odot} \right)^{-2/7} R_6^{1/7} L_{x,37}^{4/7} B_{12}^{-1/7} \mbox{esu~s}^{-1} \mbox{\ .}
\end{equation}
Figure \ref{fig:Jmax} shows $J_{\rm max}/e$ (based on Eq. \ref{eq:Jmax}, for different values of $L_{\rm x,37}$ and $B_{12}$).

\subsection{Production of the primary $\gamma-$ray photons}
\label{sect. Production of the primary gamma-ray photons}  

Relativistic protons accelerated in the gap can produce $\pi^0-$mesons
through inelastic collisions with protons of the accretion disc. The decays
of $\pi^0$ then produce $\gamma-$rays which we call primary photons.
We follow the formalism presented by \citet{Aharonian96,Aharonian00} and
adopted by \citet{Orellana07} to calculate the emissivity of $\gamma-$rays
in this case.

The pion emissivity $q_{\pi^0}(E_{\pi^0})$ can be calculated in the $\delta-$function approximation:
\begin{equation} \label{eq:qpi0}
  q_{\pi^0}(E_{\pi^0})  = n_{\rm p} \int_{E_{\rm th}}^\infty \delta(E_{\pi^0} - \kappa E_{\rm kin}) J_{\rm p}(E_{\rm p})\sigma_{\rm pp} dE_{\rm p}   =  \frac{n_{\rm p}}{\kappa}  \sigma_{\rm pp}\left(m_{\rm p}c^2 + \frac{E_{\pi^0}}{\kappa}\right) J_{\rm p}\left(m_{\rm p}c^2 + \frac{E_{\pi^0}}{\kappa}\right) \mbox{\ ,}
\end{equation}
where $n_{\rm p}$ is the density of proton targets,
$E_{\rm th}$ is the proton energy threshold for the production of $\pi^0$,
$\kappa$ ($\sim0.17$ in the broad energy range GeV to TeV) is the mean fraction of kinetic energy of the proton ($E_{\rm kin}=E_{\rm p} - m_{\rm p}c^2$)
transferred to the secondary $\pi^0$ in a collision (it includes about 6\% of contribution from the $\eta-$meson production, \citealt{Gaisser90}),
$J_{\rm p}(E_{\rm p})$ is the proton injected spectrum,
$\sigma_{\rm pp}$ is the total cross-section of inelastic $pp$ collisions, well approximated above $\sim 10$~GeV by:
\begin{equation} \label{eq:sigmapp}
  \sigma_{\rm pp} \approx 30 [0.95 + 0.06 \ln(E_{\rm kin}/\mbox{GeV})] \times 10^{-27} \mbox{\ cm}^{2}
\end{equation}
\citep{Aharonian96}.
For $J_{\rm p}$, we follow the approximation proposed by \citet{Orellana07} that protons reach the accretion disc
with the same energy gained in the gap: $E_{\rm p}\approx e\Delta V_{\rm max}$. The proton injected spectrum is thus mono-energetic:
\begin{equation} \label{eq:Jp}
 J_{\rm p}(E) = \frac{J_{\rm max}}{e \Sigma} \delta(E - E_{\rm p}) \mbox{\ } \frac{\rm protons}{{\rm s~cm}^2} \mbox{\ ,}
\end{equation}
where $\Sigma$ is the area of the accretion disc crossed by the protons.

The $\gamma-$ray emissivity due to $\pi^0$ decay is:
\begin{equation} \label{eq:qgamma}
 q_\gamma (E_\gamma) = 2 \int^\infty_{E_\pi^{\rm min}(E_\gamma)}  \frac{q_{\pi^0}(E_\gamma)}{\sqrt{E_{\pi^0}^2 - m_{\pi^0}^2c^4}} dE_{\pi^0} \mbox{\ ,}
\end{equation}
where $E_\pi^{\rm min}(E_\gamma) = E_\gamma + (m_{\pi^0}^2c^4)/(4E_\gamma)$,
$m_{\pi^0}$ is the $\pi^0$ rest mass \citep[e.g., ][]{Cavallo71}.

From Equations. \ref{eq:qgamma}, \ref{eq:qpi0}, and \ref{eq:Jp},
the rate of production of primary photons inside the disc is:
\begin{equation} \label{eq:qgamma2}
  \frac{dN_\gamma}{dt dE_\gamma} \approx \frac{2J_{\rm max} n_{\rm p}}{e} \frac{\sigma_{\rm pp}(E_{\rm p})}{\sqrt{\kappa^2(E_{\rm p} - m_{\rm p}c^2)^2 - m_{\pi^0}^2c^4}} \int_0^{2h} e^{-z/\lambda_{\rm pp}} dz  \mbox{\ \ } \frac{\rm photons}{\rm s~eV}  \mbox{\ ,}
\end{equation}
where $E_\gamma$ ranges between $E_{{\rm inf}} = 0.5E_{\pi^0}(1 - v_{\pi^0}/c)$ and $E_{\rm sup} = 0.5E_{\pi^0}(1 + v_{\pi^0}/c)$
($dN_\gamma/dtdE_\gamma=0$ outside of this range), $v_{\pi^0}$ is the velocity of the pion with energy
$E_{\pi^0} = \kappa(E_{\rm p} - m_{\rm p}c^2)$ \citep[e.g.,][]{Stecker71}, $h$ is half thickness of the accretion disc,
and $\lambda_{\rm pp}=1/(\sigma_{\rm pp}n_{\rm p})$ is the mean free path of the proton entering into the disc.

\begin{figure}
\begin{center}
  \includegraphics[width=8.5cm]{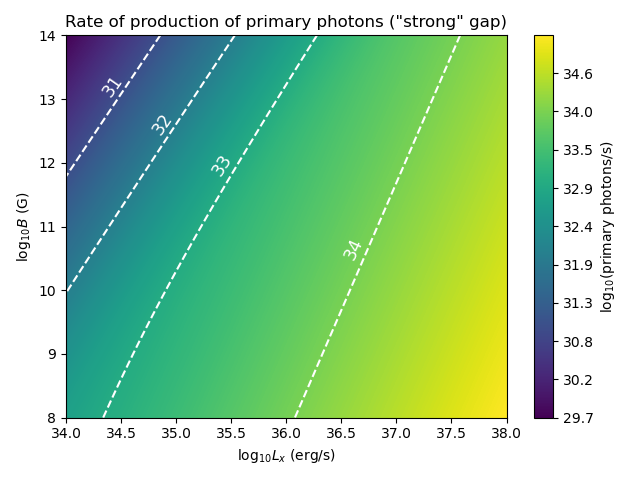}
  \includegraphics[width=8.5cm]{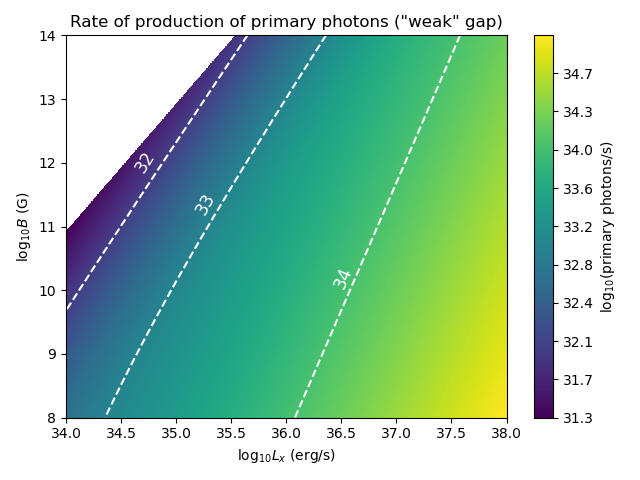}
\end{center}
\caption{Rate of production of primary photons assuming gaps for strongly (left) and weakly shielding (right) cases, as function of the X-ray luminosity and magnetic field strength (at the surface of the NS). The white region in the right panel corresponds to $E_{\rm p} < E_{\rm th}$, where the approximations used in this work are not valid. Dashed white lines show contours which have the same units used for the respective colour bars.}
\label{fig:dotNgamma/dt}
\end{figure}

\begin{figure}
  \begin{center}
  \includegraphics[width=8.5cm]{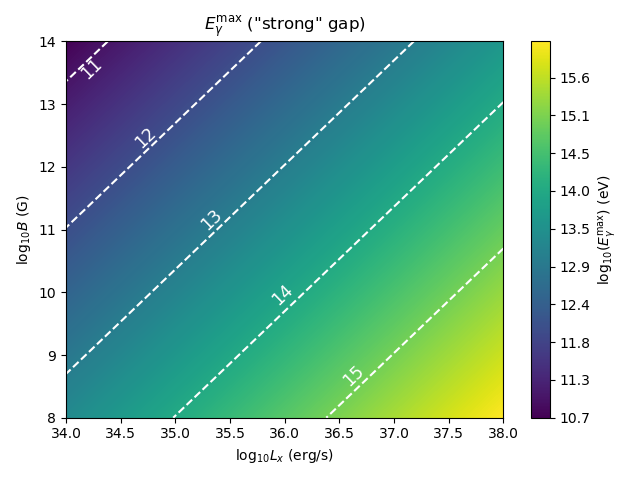}
  \includegraphics[width=8.5cm]{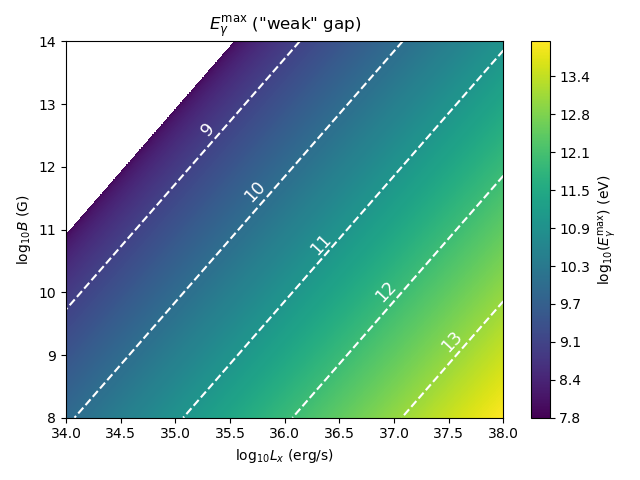}
\end{center}
\caption{Maximum energy achieved by the primary photons assuming gaps for strongly (left) and weakly shielding (right) cases, as function of the X-ray luminosity and magnetic field strength (at the surface of the NS). The white regions in the right panels have $E_{\rm p} < E_{\rm th}$, where the approximations used in this work are not valid. Dashed white lines show contours which have the same units used for the respective colour bars.}
\label{fig:E_sup/dt}
\end{figure}

Figure \ref{fig:dotNgamma/dt} shows the rate of production of primary photons (Eq. \ref{eq:qgamma2} integrated over the energy)
assuming gaps for strongly and weakly shielding cases, for different values of $L_{\rm x,37}$ and $B_{12}$.
Similarly, Fig. \ref{fig:E_sup/dt} shows the maximum energy achieved by the primary photons.

\subsection{Accretion disc structure and configuration of the magnetic field}
\label{sect. Accretion disc structure and configuration of the magnetic field}

To parametrize the thickness $h$, density $n_{\rm p}$, and temperature $T$ of the disc as a function of radius and mass accretion rate,
we adopted the analytic solution for the standard accretion disc structure presented in \citet{Vietri08},
which is a slightly modified version of that presented first by \citet{Shakura73}
and later by \citet{Treves88}, in a version better suited to astrophysical applications.

We assume a dipole magnetic field aligned with the rotational axis of the NS.
To describe the main properties of the NS magnetic field inside the accretion disc,
we consider the magnetically threaded disc (MTD) model.
Many versions of this model have been proposed (see, e.g., \citealt{Ghosh79a, Ghosh79b, Wang95} and, for a review, \citealt{Ghosh07})
and numerous analytical and numerical studies have been carried out (e.g. \citealt{Campbell87}, \citealt{Kluzniak07}, \citealt{Naso13}, \citealt{Jafari18}, \citealt{Romanova21}, and references therein).
The idea behind this model is that the magnetic field lines crossing the disc are distorted in the $\phi-$direction to form a poloidal field component,
due to the shearing motion between the differential rotation of the accretion disc and the NS spin.
For the poloidal field component, we adopted one of the configurations discussed by \citet{Wang95},
where $B_\phi$ is limited by diffuse decay due to turbulent mixing within the disc:
\begin{eqnarray}
  B_{\rm z} & = & -\eta \frac{\mu}{R^3} \label{B_z Wang95} \\
  B_\phi   & = & B_{\rm z} \frac{\gamma_{\rm disc}}{\alpha_{\rm disc}} \frac{\Omega - \Omega_{\rm K}}{\Omega_{\rm K}}  \mbox{\ ,} \label{B_phi Wang95} 
\end{eqnarray}
where $\eta \leq 1$ is a screening coefficient which is constant over a wide range of $R$, and we assumed to be equal to 0.2, according to \citet{Ghosh79a},
$\mu$ is the NS magnetic moment, $B_\phi$ is the poloidal component of the magnetic field in the upper surface of the accretion disc (in the lower surface, it has
the same magnitude and opposite direction), $\gamma_{\rm disc}\approx 1$ is a numerical factor that describes the transition in the $z$-direction
between the Keplerian motion (inside the accretion disc) and the corotation with the NS (outside of the disc)
\citep[for more details, see ][]{Ghosh79b, Wang95}, $\alpha_{\rm disc}$ is the viscosity parameter \citep[we assumed $=0.05$, according to][]{Penna13},
$\Omega$ is the angular velocity of the NS, $\Omega_{\rm K}$ is the Keplerian angular velocity of the accretion disc.
We assumed that the strength of the poloidal field component varies linearly within the disc from $B_\phi$ (upper surface) to $-B_\phi$ (lower surface).
Inside the disc, we neglected any effect of advection of the field inward (thus any additional radial component of the magnetic field),
due to the accretion of matter \citep[see, e.g.,][]{Jafari18}.

\subsection{Cascades}
\label{sect. cascades}

Some of the primary photons produced inside the accretion disc can traverse the entire disc without undergoing any interaction.
The others produce cascades of $e^\pm$ pairs and photons, until they escape from the disc.
Similarly, outside of the accretion disc, photons, electrons, and positrons can produce cascades, until they possibly reach the observer
or achieve an energy which is not relevant for our calculations ($<10$~GeV).
The processes involved in the cascades we considered are:
\begin{itemize}
\item[\emph{i)}] pair production in the Coulomb field of nucleus;
\item[\emph{ii)}] pair production in photon-photon collisions;
\item[\emph{iii)}] magnetic pair production;
\item[\emph{iv)}] bremsstrahlung;
\item[\emph{v)}] inverse Compton;
\item[\emph{vi)}] synchrotron and curvature.
\end{itemize}
  The fields of soft photons that can interact with $e^\pm$ and $\gamma-$ray photons can be roughly limited to three:
  the photons from the accretion disc, those emitted by the companion star, and the Extragalactic Background Light (EBL).
  To have a general idea on the influence of these photon fields on the development of cascades
  in an XRB, we calculated their numerical density in the vicinity of the NS.
  For the spectrum emitted by the accretion disc, we considered the standard model of
  the multi-temperature blackbody spectrum of a thin accretion disc \citep[see, e.g., ][ and references therein]{Frank02}
  and we assumed a NS with mass $M_{\rm NS}=1.4$~M$_\odot$, radius $R_{\rm NS}=12$~km,
  X-ray luminosity $L_{\rm x}=10^{37}$~erg~s$^{-1}$ and $B=10^{12}$~G (at the poles).
  For the spectrum emitted by the donor star surface, we assumed a simple blackbody model with
  $T_{\rm eff}=25000$~K, radius $R_{\rm d}=11$~R$_\odot$, and orbital separation $a=1.6\times 10^{13}$~cm
  (roughly  corresponding to $P_{\rm orb}=100$~d for a typical Be star).
  The density of photons from the accretion disc at an equatorial distance from the NS and height from the
  equatorial plane of the disc equal to $R_0$ is $\approx 10^{13}$~cm$^{-3}$, about a factor $\sim 30$
  larger than that of the photon field from the companion star.
  This factor shows substantial source-to-source variability: it is primarily affected by the properties
  of the accretion disc and the separation between the two stars.
  The influence of the photon field emitted by the donor star on the
  development of the cascades was investigated extensively by
  \citet{Bednarek97, Bednarek00}, and \citet{Sierpowska05}.
  They showed that when the radiation field from the donor star is taken into account,
  the observed $\gamma-$ray emission might show an orbital modulation.
  This is due to the anisotropic radiation field produced by the donor star through which
  the $\gamma-$ray radiation propagates, and the fact that the intensities of the
  secondary $\gamma-$rays depend on the angle of observation measured from the direction
  of propagation of the photons defined by the place of injection of the $\gamma-$rays (the NS)
  and the origin of the soft photons (the donor star).
\begin{figure}
  \begin{center}
    \includegraphics[width=7cm]{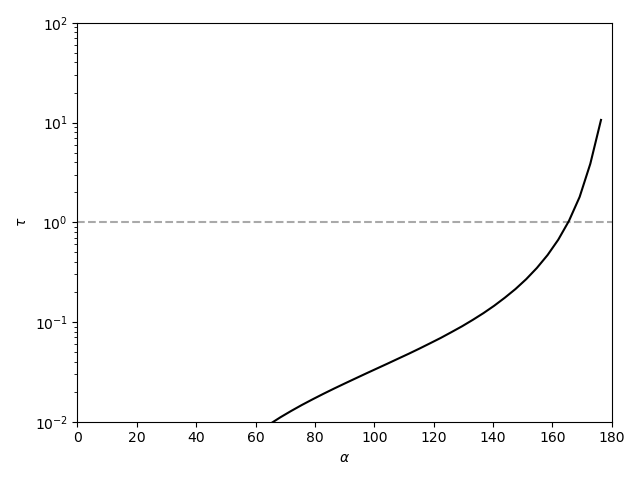}
  \end{center}
  \caption{The optical depth for a $\gamma-$ray photon with energy $E_\gamma=1$~TeV injected with an angle $\alpha$ in
    the photon field of the donor star (see also Appendix \ref{appendix:tau}).}
  \label{fig:tau_vs_alpha}
\end{figure}
To further evaluate the possible influence of the donor star radiation on the evolution of the cascade, we followed a method similar to that described in \citet{Bednarek97} to compute the optical depth for the propagation of $\gamma-$ray photons in the donor star photon field as a function of the angle of the $\gamma-$ray photon injection $\alpha$ (see Appendix \ref{appendix:tau}). Assuming that the donor star emits blackbody radiation, and assuming the same values for $T_{\rm eff}$, $R_{\rm d}$, and $a$ as above, we find that the optical depth is low over a wide range of angles $\alpha$ (Fig. \ref{fig:tau_vs_alpha}). Hence, in
  our calculations we do not take into account the effects of the donor star radiation field
  on the observed $\gamma-$ray emission, preferring
  a more streamlined approach to highlight the key aspects of the evolution of the cascades influenced by the physical processes
  more closely related to the NS itself and its immediate environment.
  On the other hand, future developments of this model will be important to incorporate the effects
  of the interactions with the photon field of the donor star, to further explore their
  potential impact on the overall $\gamma-$ray emission.
  \citet{Bednarek97} showed that these have a significant influence on the cascades for binary systems with smaller orbital separations than those considered here.
  The significantly lower density of photons in the EBL \citep[see, e.g., ][]{Cooray16} shows that they have a negligible impact on the
  development of cascades in Galactic XRBs. For instance, when considering $\gamma-$ray
  photons with energies of the order of TeV,
  the mean free path for $\gamma\gamma \rightarrow e^{+/-}$
  interactions (with EBL photons) is approximately 100~Mpc \citep[see, for example, ][]{DeAngelis13, Berezinsky16}.

Cascades are calculated in the work presented here using Monte Carlo simulations, in a framework
  based on that presented in \citet{Protheroe86}. This is one of the various
  computational and analytical techniques that have been proposed to describe the
  evolution and output of cascades in astrophysical objects. For alternative methods see,
  among the others, \citet{Aharonian03} and \citet{Berezinsky16}.
  Figure \ref{fig. flow-chart-A}
  in Appendix \ref{sect. appendix computational procedure} provides a general outline
  of the procedure used in our study to calculate the observed $\gamma-$ray emission
  and in particular to simulate the cascades.

The \emph{first step} (see Fig. \ref{fig. flow-chart-A}) of these calculations is to provide a list of input parameters to describe
  the properties of the binary system to be studied. These are: the X-ray luminosity,
  the magnetic field intensity at the polar cap,
  the distance of the binary system from the observer, the type of shielding in the accelerating gap (strong or weak),
  and the number of $N$ primary photons to simulate.

The \emph{second step} involves the acceleration of protons and the production of primary photons.
  The analytical calculations are described in Sections \ref{Sect. Engine}
  and \ref{sect. Production of the primary gamma-ray photons}.
  For each of the $N$ primary photons, the position, energy, and direction are simulated.
For a given proton accelerated in the gap, which is assumed to enter
perpendicularly in the accretion disc (due to the  ``aligned rotator'' assumption),
its radial distance from the NS, $R$, is sampled from a uniform
distribution in the range $R_0-R_{\rm A}$ \citep{Anchordoqui03}.
For a given input X-ray luminosity $L_{\rm x}$ and the radial distance $R$ just sampled,
the proton density (from the standard accretion disc structure equations),
the potential drop (Eqs. \ref{eq:DV_strong} or \ref{eq:DV_weak}),
and the cross section of inelastic $pp$ collision (Eq. \ref{eq:sigmapp})
are calculated. Then, the height $h$ in the region of the disc where the $\pi^0$ decay occurs
is determined by sampling the path of the proton from an exponential distribution
with mean free path \citep[see, e.g., ][ for the details of this type of calculations]{Protheroe86}.
The energy of the primary photon is then sampled from the energy distribution of photons from $\pi^0$ decay,
in the range $E_{{\rm inf}}-E_{{\rm sup}}$. The direction of the primary photons
($\theta$, where $\theta=0$ corresponds to the direction normal to the accretion disc plane)
is sampled using the aberration light formula in the accretion disc rest frame \citep[see, e.g., ][]{Rybicki79}.

\emph{Third step}: inside the accretion disc. Primary photons can start cascades through different processes.
For each process, the pairs or photons produced have energies and directions sampled from the respective distributions laws.
For pair production in Coulomb field nucleus we used the Davies-Bethe-Maximon formula for screened
point nucleus for extreme relativistic energy \citep{Motz69}.
For pair production in photon-photon collisions inside the accretion disc, we adopted the treatment of \citet{Protheroe86}.
For magnetic pair production we used the treatment of \citet{Erber66} and \citet{Daugherty83}.
For each photon, given the local conditions in the accretion disc
(see Sect. \ref{sect. Accretion disc structure and configuration of the magnetic field}),
the pair production process that dominates over the others is selected by comparing the
interactions lengths sampled by the characteristic distribution of each pair production process.
\begin{figure}
  \begin{center}
    \includegraphics[width=7cm]{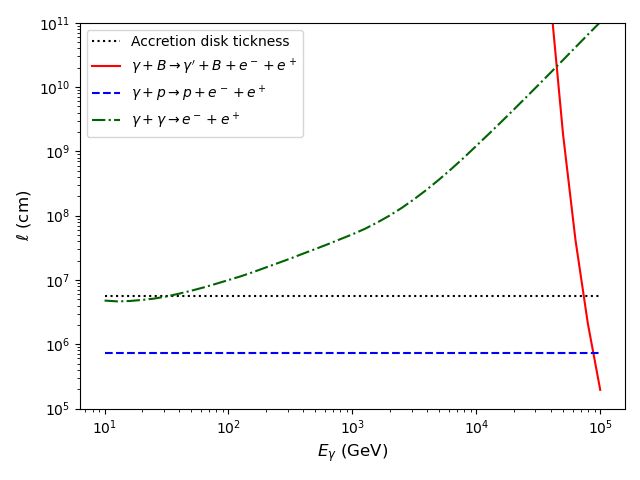}
  \end{center}
  \caption{An example of possible interaction lengths as a function of the $\gamma-$ray photon energy, inside the disc.}
  \label{fig:interaction lengths}
\end{figure}
Figure \ref{fig:interaction lengths} shows possible interaction lengths for different pair production processes
  as a function of the $\gamma-$ray photon energy, inside the disc. They are obtained assuming
  $L_{\rm x}=5\times 10^{37}$~erg~s$^{-1}$, $B=5\times 10^{12}$~G (at the polar cap), $\gamma-$ray photon
  located at a distance $R_{\rm A}$ from the NS and with a direction perpendicular to the magnetic field line.
  Figure \ref{fig:interaction lengths} has to be considered just as an example: the exact position and directions
  of the $\gamma-$ray photons, $L_{\rm x}$, and the properties of the local magnetic field
  affect the mutual values of the interaction lengths on a case-by-case basis.
Electrons and positrons can interact with the environment (protons in the disc, photons, and magnetic field) to
produce secondary photons.
For the inverse Compton scattering of soft photons produced by the accretion disc,
with interactions within the disc, we adopted the treatment of \citet{Protheroe86}.
The inverse Compton scattering and photon-photon interactions occurring outside of the accretion disc
involve the spatially varying anisotropic field of the soft photons emitted by the accretion disc.
\citet[][ and references therein]{Protheroe92} showed that in this case, the properties of the pair-Compton cascades
differ substantially from the cascade calculations based on an isotropic field of soft photons.
Outside of the accretion disc (\emph{fourth step}, see Fig. \ref{fig. flow-chart-A}), the $\gamma-$ray photons and electrons which escape from the disc
interact with the accretion disc soft photons mainly through ``tail-on'' collisions,
which influences the shape of the emerging $\gamma-$ray spectrum.
Therefore, for these processes outside of the accretion disc we followed the treatment described in \citet{Protheroe92},
that is, a modified version of the calculation method described in \citet{Protheroe86},
better suited to the varying anisotropic radiation field.
For bremsstrahlung, we used the treatment for complete screening and relativistic electrons described in \citet{Jackson75}.
For synchrotron, we adopted the formulation presented in \citet{Berestetskii82} that takes into account the quantum effects for
relativistic electrons in strong magnetic fields.
As pointed out, for example by \citet{Vietri08}, synchrotron radiation from an electron in a strong magnetic field dissipates efficiently
the component of its energy perpendicular to the magnetic field line. We therefore assumed that after the emission of synchrotron radiation,
the electrons follow the magnetic field line.
For curvature radiation, we considered the classical treatment \cite[e.g., ][]{Vietri08}
and \citet{Ochelkov80} for the curvature radius of a dipole magnetic field.
We neglected the curvature inside the disc because it depends on the complicated structure
of the magnetic field lines inside it, and its effect is in general small compared to
bremsstrahlung and synchrotron. 
Similarly to \citet{Orellana07}, by comparing the cooling times of these processes, we determined in each case which one dominates.
\begin{figure}
  \begin{center}
    \includegraphics[width=7cm]{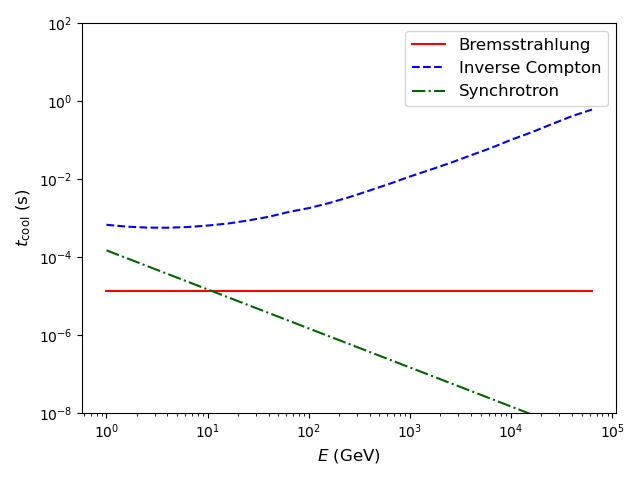}
\end{center}
\caption{An example of possible cooling times as a function of the electron energy, calculated inside the disc.}
\label{fig:t_cool comparison}
\end{figure}
Figure \ref{fig:t_cool comparison} shows an example of the cooling times of different processes as a function of the electron energy,
  calculated inside the accretion disc. We assumed $L_{\rm x}=10^{37}$~erg~s$^{-1}$,
  $B=10^{12}$~G (at the polar cap), the electron at a distance $R_{\rm A}$ from the NS
  and its trajectory perpendicular with respect to the magnetic field line.
  Different locations and directions of the electron, as well as different values of  $L_{\rm x}$ and $B$
  affect significantly the mutual values of the cooling times, which are thus calculated
  in the simulations on a case-by-case basis.
At each stage of the photon and particle production in the cascade, the new position $(R,h)$ is calculated
as described in \citet{Protheroe86}, by sampling from an exponential distribution with mean equal to the interaction length,
and subsequently by adding the the obtained value to the path travelled by cascade that originated it.
In our calculations, we allowed the cascade to continue developing until the particle or photon energy reaches energies below $10$~GeV
or the distance travelled by the cascade is lower than the distance of the source from the observer.
When the photon exceeds this distance, its energy and direction are released as output.

\section{Results}
\label{sect. results}

\begin{table*}
\begin{center}
\caption{$\gamma$-ray luminosities ($\geq 10$~GeV), for the primary photons ($L_{\rm pri}$),
  photons which escaped from the disc ($L_{\rm esc}$), and photons which reached the observer ($L_{\rm obs}$),
  assuming strong shielding, for different X-ray luminosities, and for different radiation and pair processes activated.
  We assumed $B=4\times 10^{12}$~G (i.e. a framework compatible with the Be/XRB A0535+26).}
\label{tab:a0535_expl}
\begin{tabular}{lcccc} 
  \hline
  $L_{\rm x}$       & $L_\gamma$   &                Case A$^a$    &       Case B$^b$               &        Case C$^c$              \\
  \hline
                &             &          nuclei             &           nuclei               &              nuclei            \\
                &             &                             &         + photons               &         + photons               \\
                &             &                              &                                 &          + $\vec{B}$            \\
  \hline
erg~s$^{-1}$     &             &        erg~s$^{-1}$          &          erg~s$^{-1}$           &        erg~s$^{-1}$             \\
\hline
$5\times10^{36}$  & $L_{\rm pri}$ & $6.9 \pm 0.4 \times 10^{34}$ &  $7.3 \pm 0.4 \times 10^{34}$   &   $7.0 \pm 0.4 \times 10^{34}$  \\
                 & $L_{\rm esc}$ & $5.3 \pm 0.2 \times 10^{34}$ &  $5.6 \pm 0.2 \times 10^{34}$   &   $4.4 \pm 0.2 \times 10^{34}$  \\
                 & $L_{\rm obs}$ & $5.3 \pm 0.2 \times 10^{34}$ &  $4.2 \pm 0.2 \times 10^{34}$   &   $2.5 \pm 0.2 \times 10^{34}$  \\
\hline
$10^{37}$         & $L_{\rm pri}$ & $1.76\pm 0.10\times 10^{35}$ &  $1.77\pm 0.09\times 10^{35}$   &   $1.75\pm 0.09\times 10^{35}$  \\
                 & $L_{\rm esc}$ & $9.8 \pm 0.4\times 10^{34}$  &  $9.9 \pm 0.4 \times 10^{34}$   &   $5.0 \pm 0.3 \times 10^{34}$  \\
                 & $L_{\rm obs}$ & $9.8 \pm 0.4\times 10^{34}$  &  $4.6 \pm 0.4 \times 10^{34}$   &   $1.7 \pm 0.2 \times 10^{34}$  \\

\hline
$5\times10^{37}$  & $L_{\rm pri}$ & $1.39\pm 0.06\times 10^{36}$ &  $1.37\pm 0.06\times 10^{36}$   &   $1.36\pm 0.05\times 10^{36}$  \\
                 & $L_{\rm esc}$ & $2.40\pm 0.13\times 10^{34}$ &  $2.00\pm 0.12\times 10^{34}$   &   $1.2 \pm 0.4 \times 10^{32}$  \\
                 & $L_{\rm obs}$ & $2.40\pm 0.13\times 10^{34}$ &  $5.8 \pm 2.8 \times 10^{32}$   &   $-$                          \\
\hline
\end{tabular}
\end{center}
\begin{flushleft}
  $^a$ Case A: interaction with nuclei: pair production in the Coulomb field of nucleus and bremsstrahlung.\\
  $^b$ Case B: interaction with nuclei and photons from the accretion disc: pair production in the Coulomb field of nucleus, pair production in photon-photon collisions, bremsstrahlung, inverse Compton with soft photons from the accretion disc.\\
  $^c$ Case C: interaction with nuclei and photons from the accretion disc, and with the magnetic field: pair production in the Coulomb field of nucleus, pair production in photon-photon collisions, magnetic pair production, bremsstrahlung, inverse Compton with soft photons from the accretion disc, synchrotron, and curvature.
\end{flushleft}
\end{table*}

\begin{figure*}
\begin{center}
  \begin{tabular}{lccc}
                                                                    & Case A & Case B & Case C \\
\rotatebox{90}{\hspace{1.0cm} $L_{\rm x}=5\times10^{36}$~erg~s$^{-1}$} & \includegraphics[width=5.5cm]{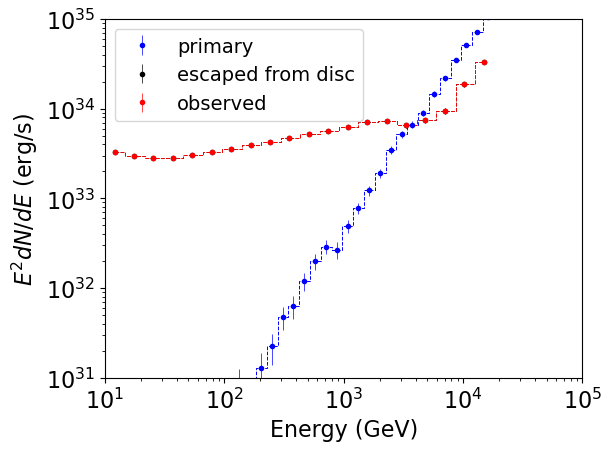} & \includegraphics[width=5.5cm]{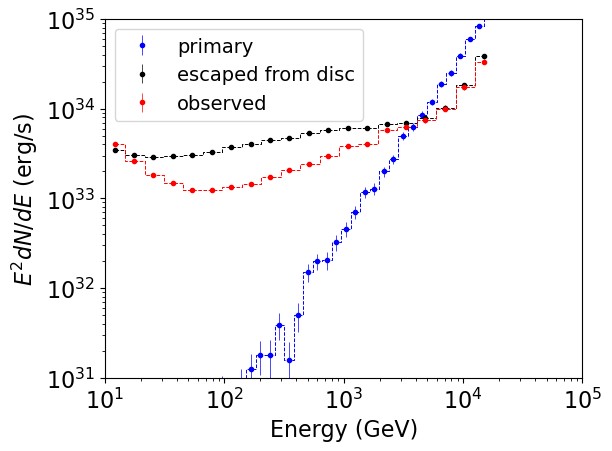} & \includegraphics[width=5.5cm]{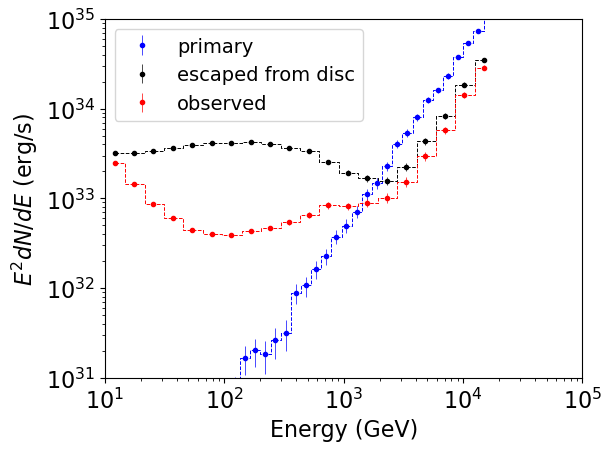}\\
\rotatebox{90}{\hspace{1.0cm} $L_{\rm x}=10^{37}$~erg~s$^{-1}$} & \includegraphics[width=5.5cm]{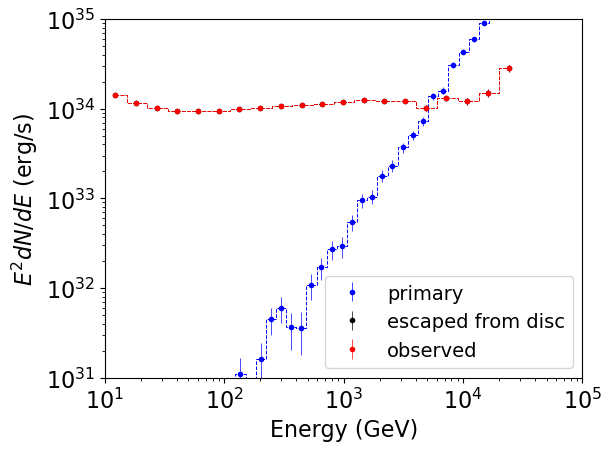} & \includegraphics[width=5.5cm]{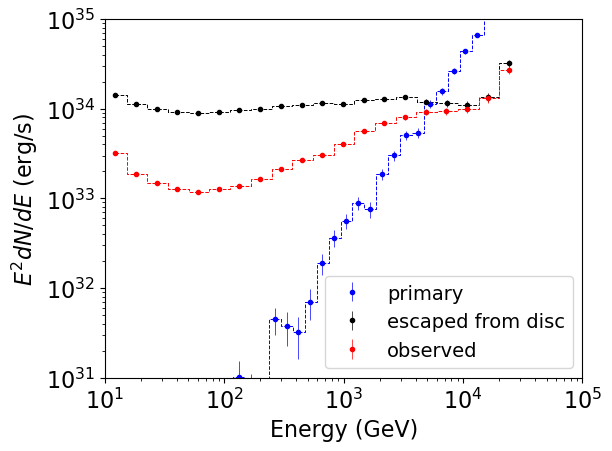} & \includegraphics[width=5.5cm]{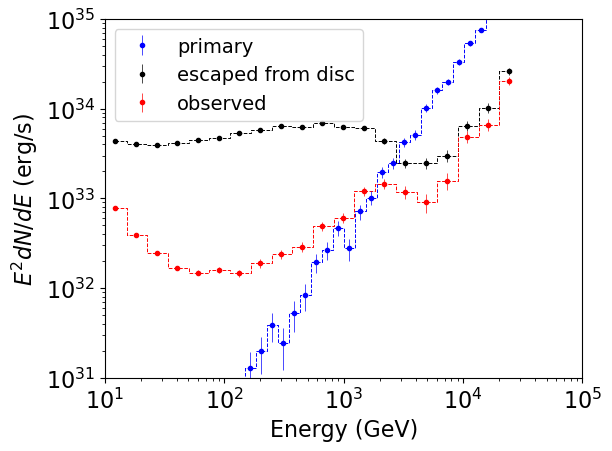}\\
\rotatebox{90}{\hspace{1.0cm} $L_{\rm x}=5\times 10^{37}$~erg~s$^{-1}$} & \includegraphics[width=5.5cm]{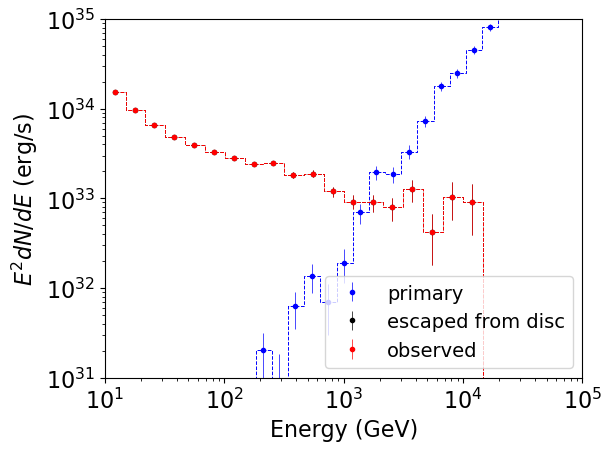} & \includegraphics[width=5.5cm]{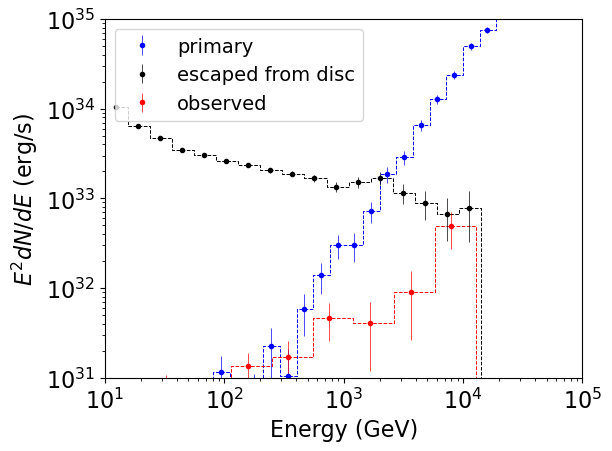} & \includegraphics[width=5.5cm]{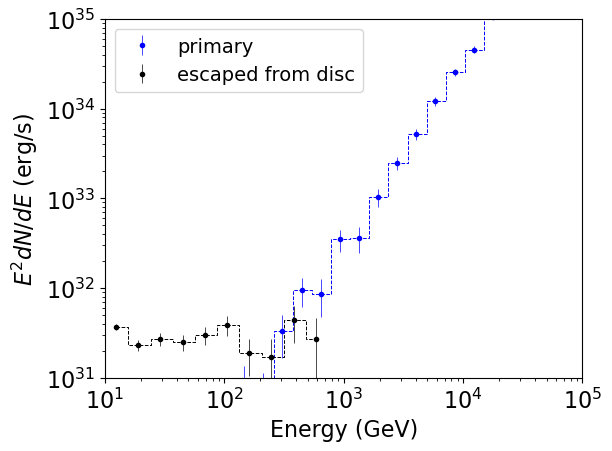}\\
\end{tabular}
\end{center}
\caption{$\gamma$-ray spectra for the primary photons,
  photons that escaped from the disc, and photons that reached the observer,
  for different X-ray luminosities, and for different radiation and pair processes activated, as described in Table \ref{tab:a0535_expl}.
  We assumed $B=4\times 10^{12}$~G and a strong shielded gap.}
\label{fig:a0535_expl}
\end{figure*}

\begin{figure*}
\begin{center}
  \begin{tabular}{lccc}
                                                                    & Case A & Case B & Case C \\
\rotatebox{90}{\hspace{1.5cm} $L_{\rm x}=5\times10^{36}$~erg~s$^{-1}$} & \includegraphics[width=5.2cm]{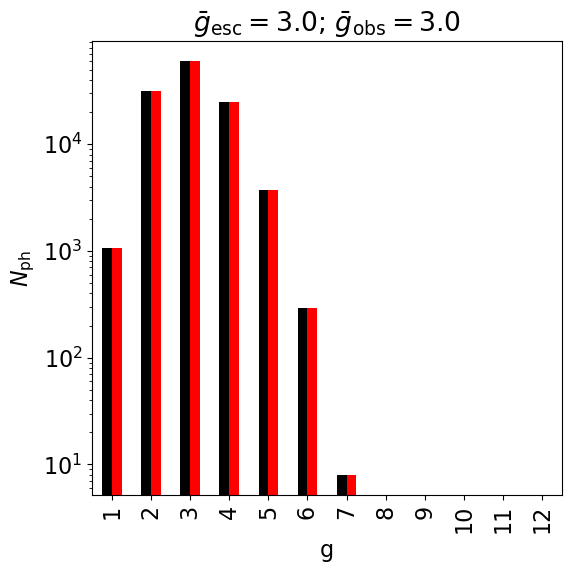} & \includegraphics[width=5.2cm]{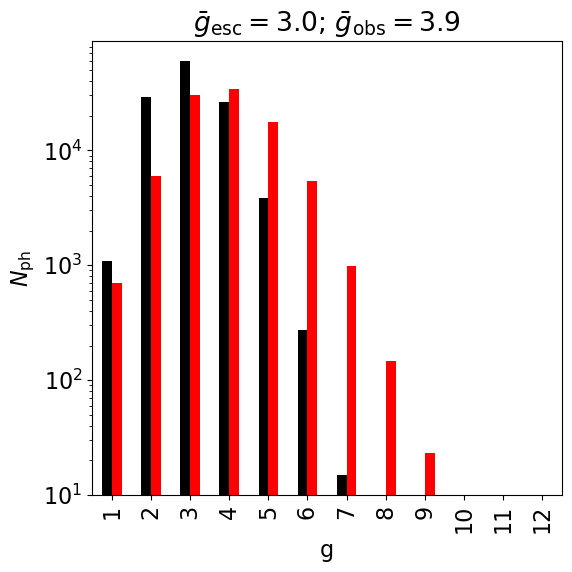} & \includegraphics[width=5.2cm]{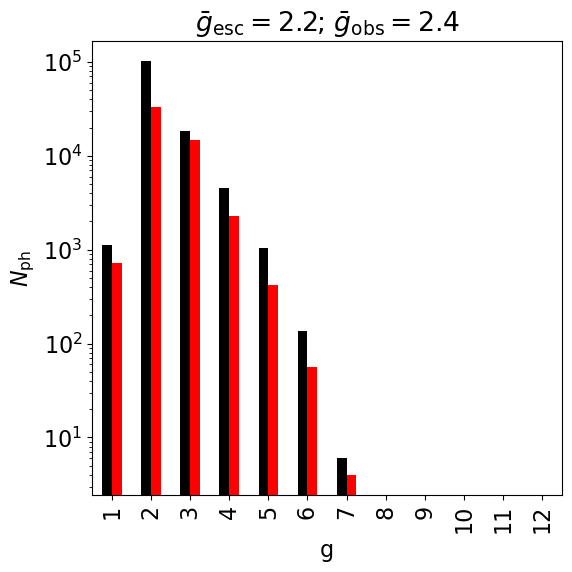}\\
\rotatebox{90}{\hspace{1.5cm} $L_{\rm x}=10^{37}$~erg~s$^{-1}$} & \includegraphics[width=5.2cm]{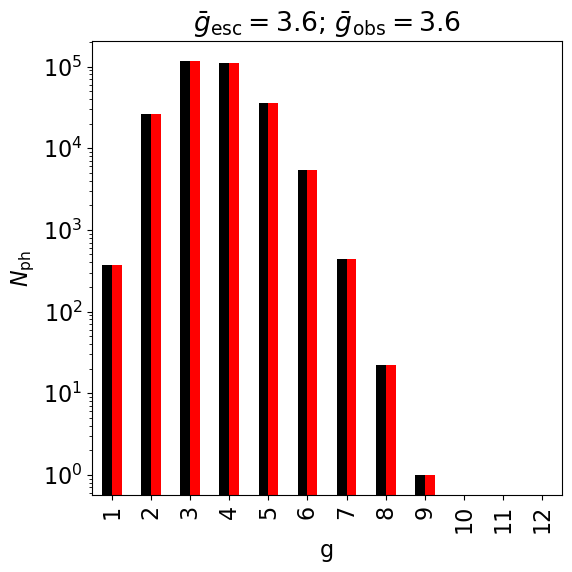} & \includegraphics[width=5.2cm]{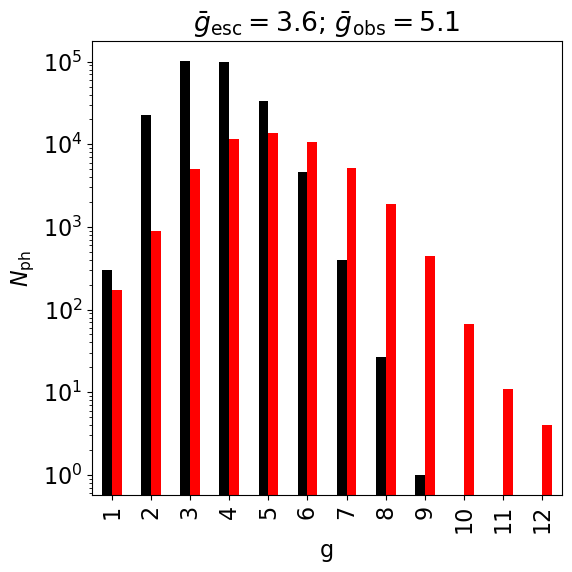} & \includegraphics[width=5.2cm]{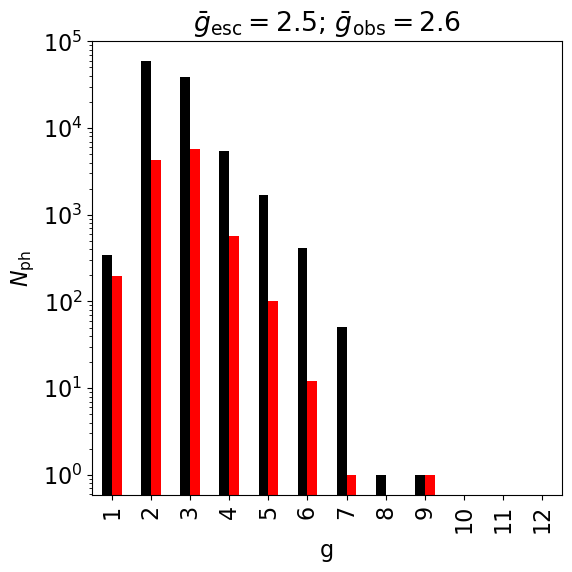}\\
\rotatebox{90}{\hspace{1.5cm} $L_{\rm x}=5\times 10^{37}$~erg~s$^{-1}$} & \includegraphics[width=5.2cm]{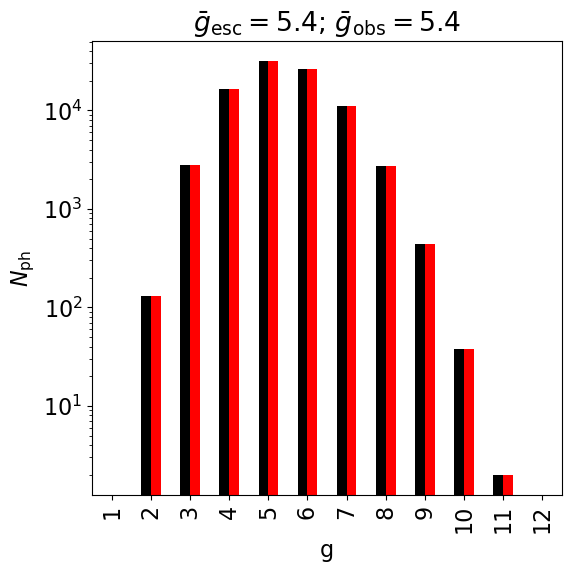} & \includegraphics[width=5.2cm]{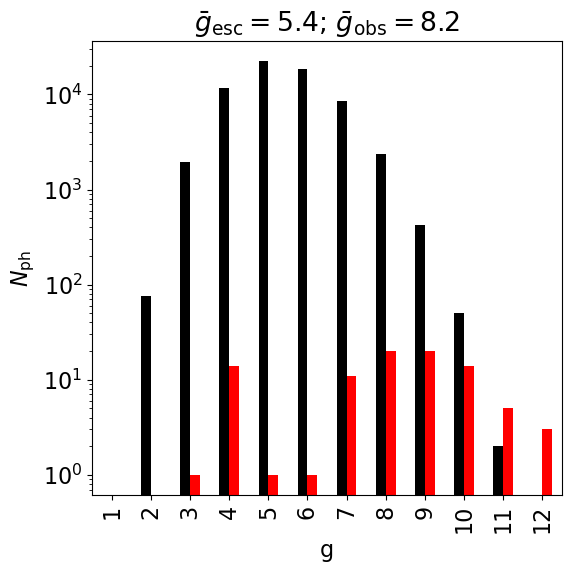} & \includegraphics[width=5.2cm]{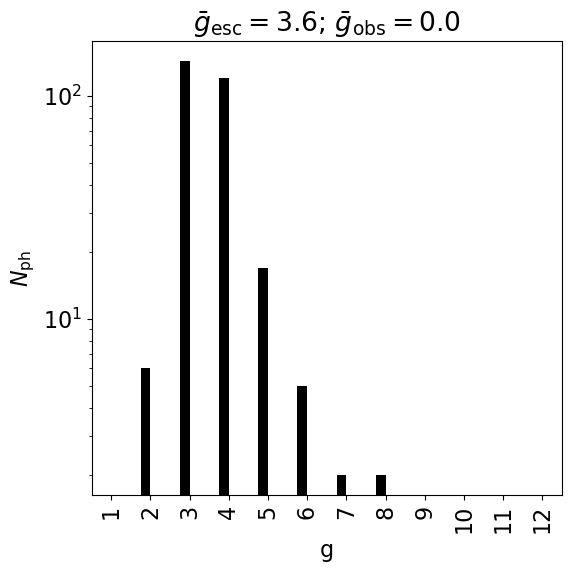}\\
\end{tabular}
\end{center}
\caption{Histograms of the generations of photons that escape from the disc (black histogram) and reach the observer (red histogram), for different X-ray luminosities
  and radiation and pair processes activated as described in Table \ref{tab:a0535_expl}.
  We assumed $B=4\times 10^{12}$~G and a strong shielded gap.
  On top of each panel is shown the average value of the generation number for photons that escaped the disc ($\bar{g}_{\rm esc}$)
  and those that reached the observer ($\bar{g}_{\rm obs}$).}
\label{fig:a0535_expl_gen}
\end{figure*}

\subsection{Contribution of different interaction processes to the $\gamma-$ray spectrum}
\label{Sect. contrib. interactions}

Table \ref{tab:a0535_expl} and Fig. \ref{fig:a0535_expl} show the observable $\gamma-$ray luminosities
  and spectra (above 10~GeV) calculated assuming a strongly shielded gap,
for three different X-ray luminosities, and for different types of interaction processes activated.
Specifically, Fig. \ref{fig:a0535_expl} shows
the $\gamma-$ray spectra from photons that escaped from the accretion disc, those observed
(i.e. photons that escaped from the disc and survived along their journey between the source and the observer
and photons that have transformed into other photons, through cascades that have occurred outside the disc),
and the spectrum of the primary photons (those created by the $\pi^0$ decay inside the disc).
Figure \ref{fig:a0535_expl_gen} shows, for each of the simulations displayed in Fig. \ref{fig:a0535_expl},
how the cascade of $\gamma-$ray photons develops inside and outside the disc. ``Gen'' (generation) 1 corresponds to primary photons,
i.e. those produced by the decay of $\pi^0$. The photons of second generation are produced by the $e^\pm$ pairs which were, in turn, produced by primary photons.
The names of the other generations of photons (gen$\geq 2$) follow the same scheme.
The results for the weakly shielded gap are in Sect. \ref{sect. appendix}.
Left panels in Fig. \ref{fig:a0535_expl} (and left columns in Table \ref{tab:a0535_expl}) show the spectra (and luminosities) obtained when only the interaction processes
involving the nuclei are activated (pair production in the Coulomb field of nucleus and bremsstrahlung).
Being the region outside of the accretion disc poor of nuclei, (indeed, they are neglected in our simulation),
in the spectra of this column the ``escaped'' spectrum coincides with the ``observed'' spectrum.
The spectra in the central column are obtained including, in addition to the interaction processes of the left column,
also those involving the interactions with the photons produced in the accretion disc
(pair production in photon-photon collisions and Inverse Compton scattering).
The right column shows spectra obtained when also the processes that involve the interaction of $e^\pm$ and $\gamma-$ray photons
with the magnetic field are considered (magnetic pair production, synchrotron, curvature).
The spectrum of primary photons is modified mostly by the interactions of photons and $e^\pm$ with the nuclei of the disc.
When $L_{\rm x}$ increases, the density of nuclei and the height of the disc in the region of interest for the $\gamma-$ray production
($R_0-R_{\rm A}$) increases, and therefore so does the number of interactions. Because of it, primary photons lose more energy and
produce more pairs and secondary photons with softer energies (Fig. \ref{fig:a0535_expl_gen}).
Observing Table \ref{tab:a0535_expl}, the peak of secondary photons produced by bremsstrahlung is reached,
for the specific case explored here, when $L_{\rm x}\approx 10^{37}$~erg~s$^{-1}$.
For higher values of $L_{\rm x}$, the spectrum becomes softer, due to the more complex cascades that develop inside the disc.
When the processes involving interactions with the soft photons from the accretion disc are activated,
the most affected spectra are those observed.
The denser is the field of soft photons emitted by the accretion disc, the larger will be the suppression of $\gamma-$ray photons
due to the production of pairs.
When also the interactions of $e^\pm$ and photons with the magnetic field are considered, the escaped and observed $\gamma-$ray radiation
are further suppressed.

\begin{figure}
\begin{center}
    \includegraphics[width=13cm]{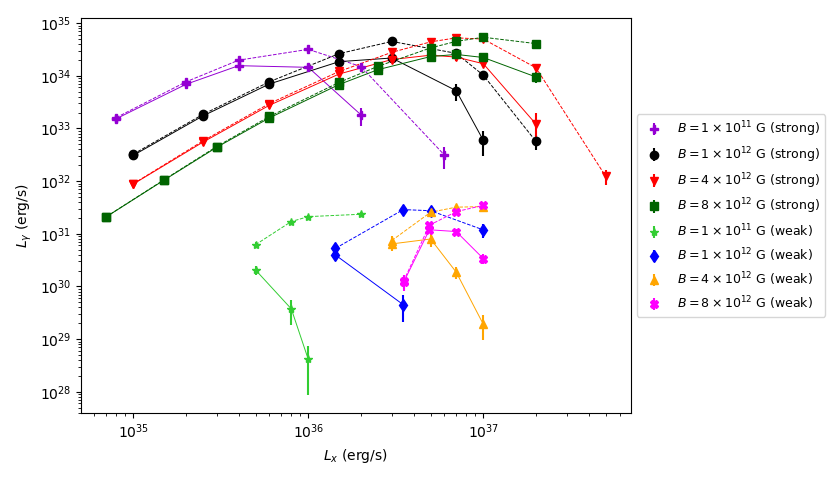}
\end{center}
\caption{$\gamma$-ray luminosities with respect to the input X-ray luminosity for different magnetic field strengths, strong and weak shielding, and for photons escaped from the disc (dashed lines) and photons that reached the observer (solid lines).}
\label{fig:Lgamma vs Lx vs B}
\end{figure}

\begin{figure}
\begin{center}
  \includegraphics[width=8.5cm]{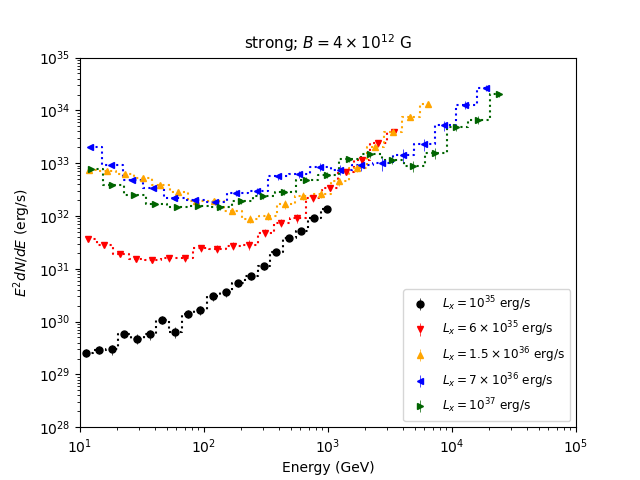}
  \includegraphics[width=8.5cm]{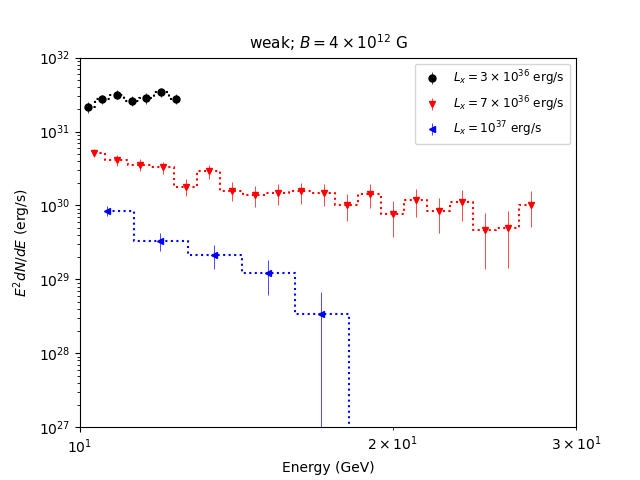}\\
  \includegraphics[width=8.5cm]{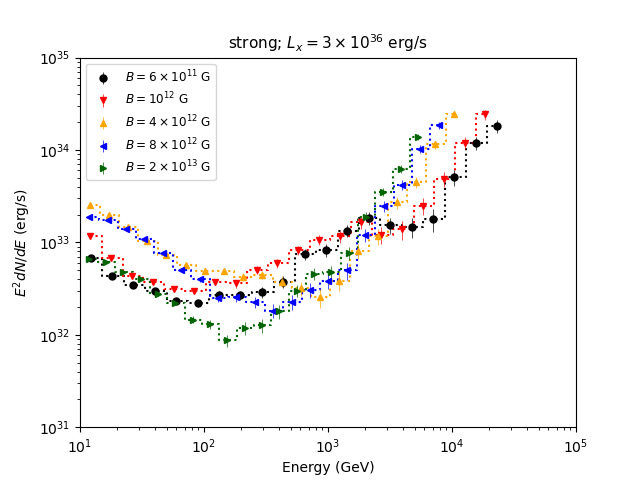}
  \includegraphics[width=8.5cm]{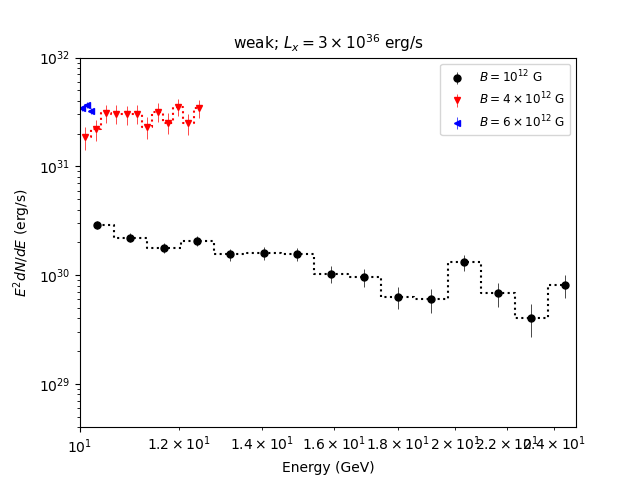}
\end{center}
\caption{$\gamma$-ray spectra of photons which reached the observer, for different luminosities (left panels) and magnetic field strenghts (right panels), assuming strong (upper panels) and weak (lower panels) shielding.}
\label{fig:general spectra}
\end{figure}

\subsection{$L_\gamma$ with respect to $L_{\rm x}$, magnetic field strength, and type of gap shielding}

We explored the properties of the $\gamma-$ray emission 
for different values of the X-ray luminosity ($L_{\rm x}$), the magnetic field strength at the poles ($B$),
and for strong and weak shielding of the gap.
The resulting $\gamma-$ray luminosities ($L_\gamma$) are shown in Fig. \ref{fig:Lgamma vs Lx vs B}, while some sample spectra
(``escaped'' from the disc and ``observed'') are shown in Fig. \ref{fig:general spectra}.
In general, $L_\gamma$ increases with $L_{\rm x}$, it reaches a peak, and then starts to decrease. This trend is seen
both in the $\gamma-$ray emission that escapes from the disc and in the one that is observed.
The rise branch of $L_\gamma$ depends mostly on the properties of the accelerating mechanism (Sect. \ref{Sect. Engine}),
which produces more protons (and consequently primary photons) when $L_{\rm x}$ increases.
The decay branch is caused by the loss of energy due to the formation of more complex cascades
that reduce the number of photons with very high energy.
The $\gamma-$ray luminosities for the weakly shielded gap are $\sim 10^3-10^4$ times lower than for strongly shielded gaps,
due to the lower energy achieved by the primary photons (Figs. \ref{fig:deltaV} and \ref{fig:E_sup/dt}).
Figure \ref{fig:Lgamma vs Lx vs B} also shows that when $B$ increases, the rise branch has a lower $L_\gamma$
and the peak is reached at higher $L_{\rm x}$.
The lower $L_\gamma$ in the rise branch as $B$ increases is due to the dependence of $J_{\rm max}$ with the magnetic field strength
(see Eq. \ref{eq:Jmax} and Fig. \ref{fig:Jmax}).
To understand the shift of the peak of $L_\gamma$ to larger $L_{\rm x}$, we have to recall that as $B$ increases, the region of the disc hit by accelerated
protons, $R_0-R_{\rm A}$, is further away from the NS. As the distance from the NS increases, $n_{\rm p}$ (inside the disc) decreases.
Therefore, for a given $L_{\rm x}$, when $B$ increases, the cascades in $R_0-R_{\rm A}$ are less developed because less interactions occur: the initial energy
of the primary photons is less distributed to $e^\pm$ and photons of higher generations and lower energies.
Therefore, when $B$ increases, it is necessary to increase $L_{\rm x}$ to push the region $R_0-R_{\rm A}$ inwards,
where $n_{\rm p}$ and the field of X-ray photons from the disc are sufficiently high to produce a more complex cascade that reduces
the flux of $\gamma-$ray photons.

Figure \ref{fig:general spectra} shows some examples of the different shapes of the observable spectra obtained from our simulations,
assuming different values of $L_{\rm x}$, $B$, and for strong and weak shielding of the gap.
The top left panel shows that, for strong shielding, the spectra are flatter as $L_{\rm x}$ increases.
This happens because the region where protons hit the disc moves inwards, where $n_{\rm p}$ is larger, and therefore more complex
cascades are produced. They redistribute the high energies of primary photons to photons having less energies.
For strong shielding, when $B$ increases (and $L_{\rm x}$ is fixed, as shown in the bottom left panel of Fig. \ref{fig:general spectra}),
there are two mechanisms working against each other. The current of accelerated protons (Eq. \ref{eq:Jmax})
and their energy (Eq. \ref{eq:DV_strong}) are inversely proportional to the magnetic field strength.
Therefore, the rising branch of the spectrum, that maps the original primary photons spectrum
(see also Fig. \ref{fig:a0535_expl}) moves to lower energies.
On the other hand, when $B$ increases, the region where protons hit the accretion disc moves outwards,
where $n_{\rm p}$ is lower. Therefore, the resulting cascades are less developed and more primary photons
(and less photons of secondary or higher generation) are able to escape from the disc,
leading to an increase of the observed flux above $\sim 1$~TeV.
This leads to an increase in the slope of the branch.
The spectra obtained for the weak shielding case are much softer than those for strong shielding case, again due to the 
lower energy achieved by the primary photons in the weak shielding case (Figs. \ref{fig:deltaV} and \ref{fig:E_sup/dt}).
Simulations with magnetic fields of $10^{9}$~G  and $10^{10}$~G did not provide any observable photons,
due to the higher $n_{\rm p}$ in the region $R_0-R_{\rm A}$ of the disc.

\subsection{Beaming factor}

\begin{figure}
\begin{center}
  \includegraphics[width=8.5cm]{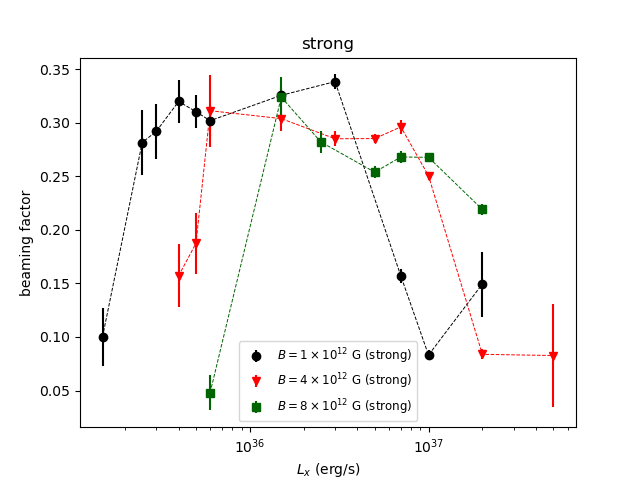}
  \includegraphics[width=8.5cm]{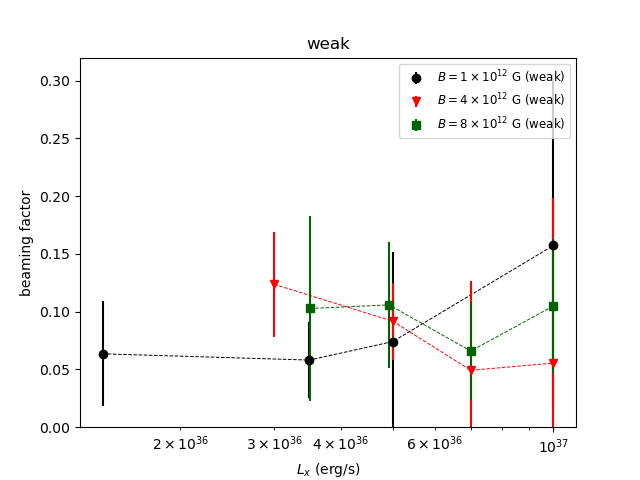}
\end{center}
\caption{Beaming factor as function of the X-ray luminosity, for three different values of the magnetic field strength,
  assuming strong (left panel) and weak (right panel) shielding.}
\label{fig:beaming 1}
\end{figure}

For a bipolar emission, the beaming factor is given by $f_{\rm b}\approx 1 - \cos\theta$,
where $\theta$ is the half-opening angle (with respect to the normal of the accretion disc plane). We calculated, through our simulations, $f_{\rm b}$ for different values of
the X-ray luminosity and magnetic field strength, assuming strongly and weakly shielded gap and a cone-beam that contains 90\% of the total emitted
$\gamma-$ray radiation. The results are shown in Fig. \ref{fig:beaming 1}.
The panel for strong shielding in the figure shows that $f_{\rm b}$ increases with $L_{\rm x}$, until it reaches a maximum.
The point where $f_{\rm b}$ starts to decrease occurs at higher X-ray luminosities as the magnetic field increases.
These trends can be explained as follows.
When $L_{\rm x}$ is low ($\lesssim 10^{35-36}$~erg~s$^{-1}$), the density and height of the disc in the region where the
accelerated protons hit the disc ($\sim R_0-R_{\rm A}$) are relatively small. Most of the primary photons escapes from the disc
without further interactions. Therefore, the beaming factor is small (nearly zero at very low $L_{\rm x}$, due to the initial
assumptions of aligned rotator and that protons enter perpendicularly in the accretion disc).
When $L_{\rm x}$ increases, the primary photons interact more inside the accretion disc, first producing $e^\pm$ pairs,
then secondary photons (and, eventually, higher generations).
A substantial fraction of $e^\pm$ starts to be bended by the poloidal magnetic field inside the disc,
and the directions of the secondary photons plus the primary photons that escaped from the disc will thus cover a wider angle.
When $L_{\rm x}$ increases further, the density and thickness of the disc increases such that a more complex cascade
develops inside the disc. The $e^\pm$ pairs that are created at each step of the cascade will more closely follow the direction
of the magnetic field lines inside the disc. The emerging beam of radiation will thus start to become narrower again.

For a given $L_{\rm x}$, the region where protons hit the disc ($\sim R_0-R_{\rm A}$) moves outwards as the magnetic field strength increases.
Moving towards the outermost regions, $n_{\rm p}$ decreases.
Therefore, for $L_{\rm x}\gtrsim 5\times 10^{36}$~erg~s$^{-1}$,
when the magnetic field strength increases, the cascades inside the disc are
less developed and for the above reasons, $f_{\rm b}$ increases.
The right panel of Fig. \ref{fig:beaming 1}, which shows for the weak shielding case, covers a smaller range of X-ray luminosities
compared to the left panel. We have shown previously this behaviour (see, e.g., Fig. \ref{fig:Lgamma vs Lx vs B}).
In the weak shielding case, indeed, the $\gamma-$ray emission is much lower than in the strong case.
In this case, the ability to simulate $\gamma-$ray photons for low and high X-ray luminosities (i.e., when the source is
expected to be very faint in the $\gamma-$ray band) is highly reduced,
since extremely computationally demanding simulations would be required.
The right panel of Fig.  \ref{fig:beaming 1} shows that $f_{\rm b}$ does not show a significant variability within the uncertainties.
It shows also an overall lower $f_{\rm b}$ than in the case of strong shielding.
This can be explained with the lower number of interactions in the weak shielding scenario (see Fig. \ref{fig:a0535_expl_gen_we}, compared to Fig. \ref{fig:a0535_expl_gen}),
which leads to photons with smaller deviations from the original trajectory of the primary photons (which are almost perpendicular to the disc plane).

\section{Discussion}

In Sect. \ref{sect. results} we showed that, in the framework of the model presented in this work, HMXBs,
which typically have magnetic fields of $B\approx 10^{12}-10^{13}$~G, are potentially bright $\gamma-$ray sources.
High X-ray luminosities and low magnetic fields favour the production of high rates of protons with high energy
(see Figures \ref{fig:deltaV}, \ref{fig:Jmax}, \ref{fig:dotNgamma/dt}, \ref{fig:E_sup/dt}).
Therefore, NS-LMXBs, which have typically low magnetic field strength ($B\approx 10^8-10^{10}$~G), observed during
outbursts ($L_{\rm x}\approx 10^{36}-10^{37}$~erg~s$^{-1}$), could also be bright $\gamma-$ray sources.
On the other hand, when a NS with low magnetic field is bright in X-rays, the density in the $R_0-R_{\rm A}$ region of the accretion disc is larger
and reduces most of the escaping $\gamma-$ray photons.
In this regard, we performed simulations assuming $L_{\rm x}=3\times 10^{36}$~erg~s$^{-1}$, for strong and weak shielding for the gap,
and $B=10^9$~G and $B=10^{10}$~G which did not produce any emission of $\gamma-$ray photons able to escape from the accretion disc.
\citet{Harvey22b} pointed out that, although the population of LMXBs in our Galaxy is almost
twice than that of HMXBs, its members are only half as many in the latest \emph{Fermi} catalogue, and consists of only four sources.
In addition, many LMXBs are contained in globular clusters, which are too crowded to allow a simple identification.
For these reasons, hereon we focus on HMXBs.

\subsection{Observability of HMXBs in the $\gamma-$ray band}

We assess the observability of HMXBs above 10~GeV assuming that the model developed in Sect. \ref{Sect. Calculations} is the only one possible for the production of
$\gamma-$rays in HMXBs. Given the number of HMXBs currently known in our Galaxy,
we calculate the fraction of them that could be observable by \emph{Fermi}/LAT and the Cherenkov Telescope Array Observatory (CTAO). We consider the \emph{Fermi}/LAT sensitivity in the $\sim$10$-$100~GeV band (after 4 years of survey) 
and that expected for CTAs in the $\sim0.1-100$~TeV band (after 50 hours of observations\footnote{See: \url{https://www.cta-observatory.org/science/ctao-performance/}
and \url{https://fermi.gsfc.nasa.gov/ssc/data/analysis/documentation/Cicerone/Cicerone_LAT_IRFs/LAT_sensitivity.html}}).
To date, we know in our Galaxy $\sim 54$ Be/XRBs and 32 HMXBs with OB supergiant stars \citep{Neumann23, Fortin23, Arnason21, Liu06}.
For some Be/XRBs, there are indications that during X-ray outbursts, an accretion disc forms around the pulsar
(e.g. from the observation of QPOs, from the properties of the spin period variability as function of the mass accretion rate,
and from the detection of double peaked He$_{\rm II} \lambda4686 $ emission lines; see, e.g., \citealt{Hayasaki04}, \citealt{Motch91}, \citealt{Rajoelimanana17}, and references therein).
For HMXBs with supergiant companion, instead, the presence of an accretion disc is less certain.
In general, it seems that in most of these systems the accretion is spherically symmetrical, likely not mediated
by a standard accretion disc (see., e.g., \citealt{Davidson73}).
Nonetheless, \citet{ElMellah19} showed that in some of these systems, a disc-like structure may form around the NS.
Since the model we have developed requires the presence of a standard accretion disc, we restrict to the Be/XRBs for the observability calculations.
In Sect. \ref{sect. results} we showed that the maximum $\gamma-$ray luminosity
($L_\gamma \approx 10^{33}-10^{34}$~erg~s$^{-1}$) is reached when the X-ray luminosity is $L_{\rm x} \approx 10^{36}-10^{37}$~erg~s$^{-1}$.
This X-ray luminosity is observed in Be/XRBs during the rise or decay time of a giant outburst, or during normal outbursts \citep[see, e.g., ][]{Reig11}.
The sensitivity of a 50-hours observation by CTA in the energy band $\sim0.1-100$~TeV is $F_{\rm th}^{\rm CTA}\approx 2\times 10^{-13}$~erg~cm$^{-2}$~s$^{-1}$.
For \emph{Fermi}/LAT, the sensitivity threshold in the energy band $\sim10-100$~GeV (for $\sim 1$ year of observations) is
$F_{\rm th}^{\rm Fermi}\approx 2\times 10^{-12}$~erg~cm$^{-2}$~s$^{-1}$.
Therefore, the maximum distances that a Be/XRB can have to be detected by CTA and \emph{Fermi}/LAT
are $d_{\rm th}^{\rm CTA}=\sqrt{L_\gamma/(4\pi F_{\rm th}^{\rm CTA})}\approx 6.5$~kpc
and $d_{\rm th}^{\rm Fermi}\approx 2$~kpc. In both cases, we assumed a conservative $L_\gamma \approx 10^{33}$~erg~s$^{-1}$.
The fraction of Be/XRBs within these distances are $f^{\rm Be/XRB}_{\rm 6.5\,kpc}\approx 0.7$ and
$f^{\rm Be/XRB}_{\rm 2\,kpc}\approx 0.2$.
The fraction of Be/XRBs that have shown X-ray luminosities of $L_{\rm x} \approx 10^{36}-10^{37}$~erg~s$^{-1}$ during transient events are $f_{\rm t}\approx 0.6$.
The beaming effect can greatly reduce the number of detectable XRBs in $\gamma-$rays,
enabling the detection only of those that emit the radiation close to the line of sight.
Assuming $f_{\rm b}\approx 0.3$ (i.e. roughly the beaming factor expected for $L_{\rm x} \approx 10^{36}-10^{37}$~erg~s$^{-1}$, see Fig. \ref{fig:beaming 1}),
we have, for the energy band $\sim0.1-100$~TeV: $N_{\rm obs}^{\rm CTA}\approx N_{\rm Be/XRB}f_{\rm t}f^{\rm Be/XRB}_{\rm 6.5\,kpc}f_{\rm b}\approx 7$,
while for the energy band $\sim10-100$~GeV $N_{\rm obs}^{Fermi}\approx 2$.
Considering an average outburst rate for each Be/XRB of $\approx 0.5$~outburst/year\footnote{Based on the number of outbursts of Be/XRBs detected by \emph{Fermi}/GBM
  since 2008 and reported in \url{https://gammaray.nsstc.nasa.gov/gbm/science/pulsars.html}}, a few tens of $\gamma-$ray transient events could be observed by CTAO in the next
10 years.
If the gap is strongly shielded, most of the $\gamma-$ray flux is emitted in the TeV band, while for the
weak shielding case, most of the $\gamma-$ray flux is emitted in the $\sim 10-100$~GeV band.
Therefore, $N_{\rm obs}^{\rm CTA}$ and $N_{\rm obs}^{Fermi}$ can be seen as the number of observable Be/XRBs
in the case of gaps strongly or weakly shielded, respectively.

Recently, the detections of some $\gamma-$ray counterpart candidates of HMXBs
with \emph{Fermi}/LAT have been reported \citep[][ and Table \ref{tab:list_sources}]{Harvey22, Xing19, Romero01, Li12}.
For most of these sources there is no firm detection,
or the fundamental parameters of the binary system are unknown
(especially the magnetic field strength and the spin period of the pulsar, which are relevant for the
mechanism investigated in this paper).
We selected two of these sources to perform some tailored simulations. These sources
are A0535+26 and GRO~J1008$-$57, two Be/XRBs with known magnetic field strength and spin period.
A0535+26 and GRO~J1008$-$57 shows long periods ($\sim$years) of low X-ray luminosity states ($L_{\rm x}\approx 10^{34}-10^{35}$~erg~s$^{-1}$
sporadically interrupted by bright X-ray outbursts with a duration of several days to several weeks,
and peak luminosities of $L_{\rm x}\approx 10^{38}$~erg~s$^{-1}$.
For these two sources there are also some previous observational studies in the $\gamma-$ray band, as described below.

\subsection{A0535+26}

A0535+26 is located at a distance of $\sim 1.8$~kpc \citep{Bailer-Jones21}.
The pulsar has a spin period of $\sim 103$~s \citep{Rosenberg75}
and orbits around the companion star, an O9.7-B0~IIIe type star in $\sim 111$ days \citep{Coe06, Nagase82}.
The magnetic field strength at the poles is rather well known, being measured from the detection of a cyclotron resonance scattering
feature (CRSF), with the fundamental line centered at $\sim 45$~keV \citep{Kendziorra94}.
This energy corresponds to a magnetic field of about $4\times 10^{12}$~G (see, e.g., \citealt{Revnivtsev15}).
During the 1994 and 1999 giant outbursts, A0535+26 showed broad quasi-periodic oscillations which are considered good indicators
of the presence of an accretion disc (see \citealt{Finger96, Camero-Arranz12} and references therein) during the outbursts.

Transient $\gamma-$ray radiation from a region spatially coincident with A0535+26, on a timescale of months,
was detected by EGRET (source name: 3EG~J0542+2610; energy band: 30~MeV$-$10~GeV; \citealt{Romero01, Torres01}).
\citet{Romero01} discussed the possible counterparts of 3EG~J0542+2610 at other wavelengths
and then suggested that the only known source within 95\% confidence region of the EGRET source
able to produce the observed variable $\gamma-$ray emission is A0535+26.
The $\gamma-$ray flux measured with EGRET was $(14.7 \pm 3.2)\times 10^{-8}$~ph~cm$^{-2}$~s$^{-1}$ ($>100$~MeV; \citealt{Romero01}).
Recently, \citet{Harvey22} reported a weak correlation between the X-ray and $\gamma-$ray fluxes of A0535+26
using 12.5 years of \emph{Fermi}/LAT data (energy band: 0.1$-$500~GeV).
They measured $\gamma-$ray emission from A0535+26 (with a significance in the range $2 < z < 3\sigma$)
in two bins of the lightcurve, having each time intervals of six months of \emph{Fermi}/LAT data
($z$ is a statistic measure of the significance of a $\gamma-$ray source detection based on the likelihood ratio test; see \citealt{Harvey22} and references therein).
These bins roughly correspond to the brightest X-ray outbursts
displayed by the source in December 2009 and November 2020.
The analysis of \emph{Fermi}/LAT data showed also that most of the $\gamma-$ray flux is concentrated
in the orbital phase bin $0.9\leq \phi < 1$, that preceeds the periastron (significance: $\sim 3.5\sigma$).
The \emph{Fermi}/LAT flux is $\sim 1.45\times 10^{-6}$~MeV~cm$^{-2}$~s$^{-1}$ (0.1$-$500~GeV).
The statistic was too low to extract a spectrum.
On the other hand, in a similar data analysis presented by \citet{Hou23} and based on about 13 years
  of \emph{Fermi}/LAT observations, A0535+26 is not detected. \citet{Hou23} suggest that the 
  difference between their findings and those reported by \citet{Harvey22} might be due to the
  different size of the fitting regions adopted, as well as the treatment of the spectral index, which was fixed to different values in \citet{Hou23}
  while in \citet{Harvey22} it was left free to vary.
VERITAS observed A0535+26 at VHEs during the 2009 and 2020 giant X-ray outbursts. In both observational campaigns,
the source was not detected.
During the 2009 outburst, the flux upper-limit (at the 99\% confidence level) was
$F_{\rm u.l.}=0.5\times 10^{-12}$~ph~cm$^{-2}$~s$^{-1}$ above 0.3~TeV, assuming a power law source spectrum
with photon index $\Gamma=2.5$ \citep{Acciari11}.
During the 2020 outburst, the preliminary result about the flux upper-limit (99\% c.l.) was
$F_{\rm u.l.}=2\times 10^{-12}$~ph~cm$^{-2}$~s$^{-1}$ above 0.15~TeV, assuming a power law source spectrum
with photon index in the range $\Gamma=3-4$ \citep{Lundy21}.
For the outburst of 2009, \citet{Acciari11} pointed out that the VERITAS upper-limit
is close to the $\gamma-$ray flux derived by \citet{Orellana07} by applying the \citet{Cheng89} mechanism to A0535+26.

\begin{figure}
\begin{center}
  \includegraphics[width=8.5cm]{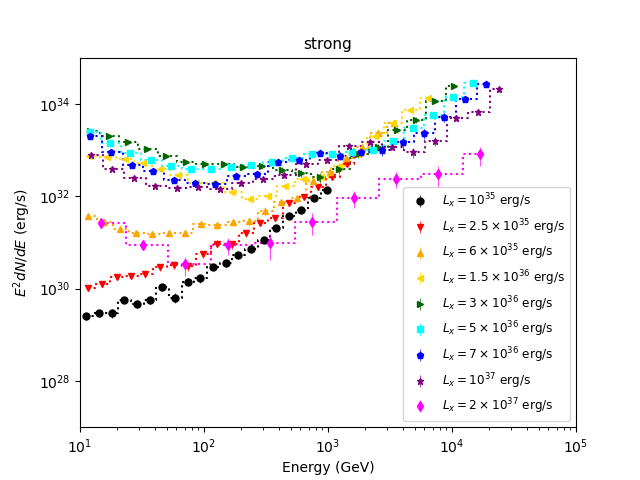}
  \includegraphics[width=8.5cm]{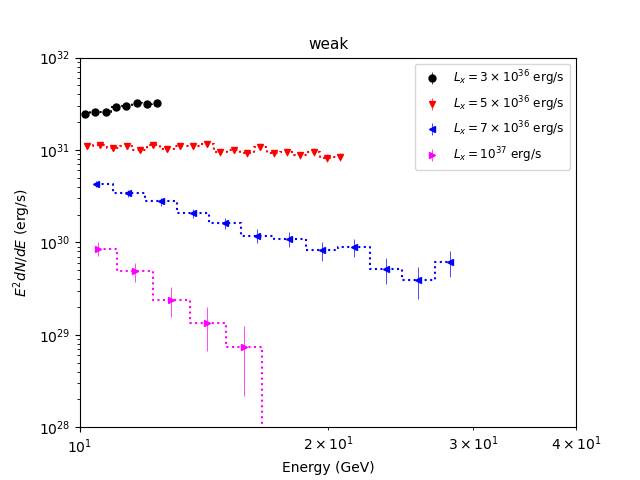}
\end{center}
\caption{$\gamma$-ray spectra of A0535+26 for the photons which reached the observer, assuming strong (left panel) and weak (right panel) shielding.}
\label{fig:A0535 spectra}
\end{figure}

We simulated the $\gamma-$ray spectra of A0535+26 for different input values of $L_{\rm x}$ (that varies from $L_{\rm x}=10^{35}$~erg~s$^{-1}$ to $L_{\rm x}\approx 2\times 10^{37}$~erg~s$^{-1}$),
assuming strong and weak shielding of the gap. The spectra are shown in Fig. \ref{fig:A0535 spectra}.
These spectra can be used as a grid for comparisons with future observations.

\begin{figure}
\begin{center}
    \includegraphics[width=13cm]{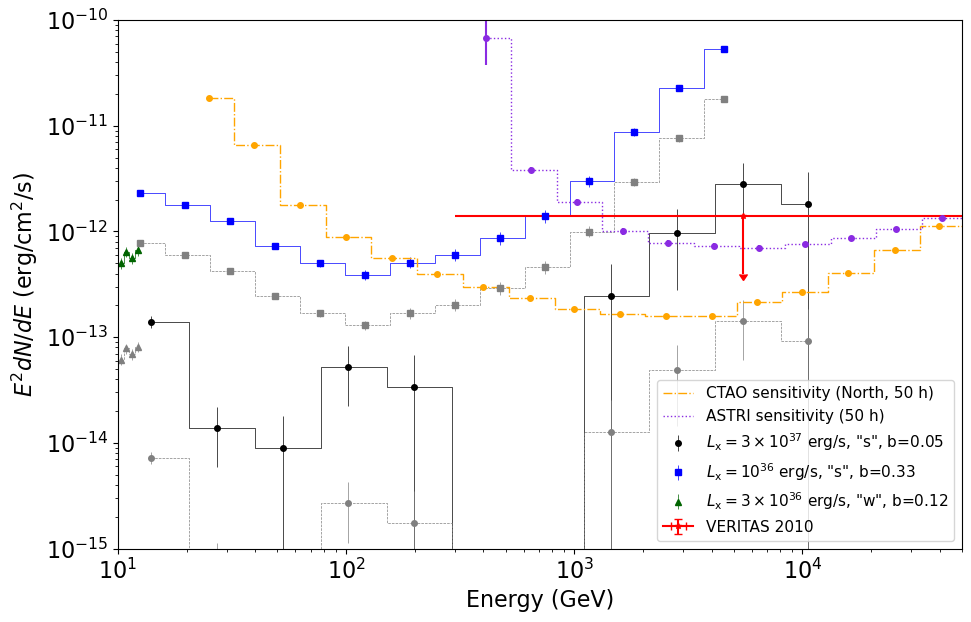}
\end{center}
\caption{Some examples of spectral energy distribitutions of A0535+26 which take into account the beaming effect, compared with the VERITAS upper-limit
  measured during the 2009 outburst and apoastron. In gray, the SEDs for no beaming are shown. Orange and blue-violet points show the
  differential flux sensitivity of the CTAO North and ASTRI Mini-Array, respectively, both for 50 hours of observations.}
\label{fig:a0535 flux spectra}
\end{figure}

To allow a direct comparison with the available VERITAS upper-limit of the outburst of 2009,
we show in Fig. \ref{fig:a0535 flux spectra} some simulated flux spectra,
which take into account the effects of the beaming.
The X-ray luminosities of the spectra obtained assuming strong shielding (``s'' in the legend of the figure)
correspond to the typical average X-ray luminosity during a giant X-ray outburst of A0535+26 ($L_{\rm x}\approx 3\times 10^{37}$~erg~s$^{-1}$)
and to the X-ray luminosity measured by \citet{Acciari11} in the $\sim50$ days following the end of the outburst, when the source had an
X-ray luminosity of the order of $L_{\rm x}\approx 10^{36}$~erg~s$^{-1}$. In both time intervals, \citet{Acciari11} reported the upper-limits
for the 0.3$-$100~TeV flux, which is the same for both X-ray luminosity states and it is shown in the figure.
Figure \ref{fig:a0535 flux spectra} also shows a spectrum obtained assuming weak shielding (``w'' in the legend of the figure),
for a luminosity state of  $L_{\rm x}\approx 3\times 10^{36}$~erg~s$^{-1}$. This spectrum is roughly comparable to the \emph{Fermi}/LAT
upper-limit obtained by \citet{Hou23} and with the marginal detection obtained by \citet{Harvey22} by averaging about 12 months of data around
two giant outbursts ($\sim 10^{-12}$~erg~cm$^{-2}$~s$^{-1}$ in 10$-$500~GeV, assuming a power law slope $\Gamma=2$ for the energy conversion).
If we assume $L_{\rm x}\approx 3\times 10^{37}$~erg~s$^{-1}$ (s), the $\gamma-$ray spectrum is in first approximation compatible with the VERITAS upper-limit,
while the $\gamma-$ray emission obtained assuming $L_{\rm x}\approx 10^{36}$~erg~s$^{-1}$ (s) is significantly larger than the upper-limit.
On the other hand, the soft part of the same spectrum, as well as the spectrum obtained assuming weak shielding are roughly consistent with
the \emph{Fermi}/LAT detection. The discrepancy between the simulated spectrum at $L_{\rm x}\approx 10^{36}$~erg~s$^{-1}$ (s)
and the VERITAS upper-limit might suggest that the accelerating gap of this source is weakly shielded.
Alternatively, if the gap is strongly shielded, other physical processes could be considered for a possible significant impact
on the emerging $\gamma-$ray emission, especially for $L_{\rm x}\approx 10^{36}$~erg~s$^{-1}$, for $\gamma-$ray emission
to fit better with the VERITAS upper-limit. For example, the effects of the UV-optical photons of the donor star might be significant;
this possibility was considered by \citet{Bednarek93, Bednarek97, Bednarek00} and it is not taken into account in our current simulations (see Sect. \ref{sect. cascades}).
Another parameter that can affect the $\gamma-$ray spectrum is the outer radius of the ring where the protons accerelareted in the gap
hit the accretion disc. We assumed $R_{\rm out}=R_{\rm A}$ \citep{Anchordoqui03}, but it is not unreasonable to assume a value of $R_{\rm out}$
slightly smaller or larger than the Alfv\'en radius. Another assumption of our simulations that might influence the result is that of aligned rotator.
Figure \ref{fig:a0535 flux spectra} also shows the differential flux sensitivities of the CTAO North 
and the ``Astrofisica con Specchi a Tecnologia Replicante Italiana'' (ASTRI) Mini-Array
\citep{Scuderi22}\footnote{CTAO data retrieved from \url{https://www.cta-observatory.org/science/ctao-performance/}. ASTRI Mini-Array data retrieved from figure 4 in \citealt{Vercellone22}.}, both for 50 hours of observations.
Future observations of A0535+26 with CTAO and ASTRI Mini-Array will have the adequate sensitivity
to test the proposed model and to identify the best ways to improve it.

\subsection{GRO J1008$-$57}

GRO J1008$-$57 hosts a pulsar with spin period of $\sim 93.6$~s \citep{Stollberg93,Wilson94}
with an orbital period of $247.8\pm0.4$~d \citep{Levine06,Coe07} and a B1-B2~Ve type donor star \citep{Coe07}.
The magnetic field strength at the poles is one of the highest ever measured
through the detection of a CRSF.
For this source, there are different detections of CRSFs in the energy range 78$-$90~keV, which
correspond to a magnetic field of about $7-8\times 10^{12}$~G \citep{Grove95, Bellm14, Ge20}.
Correlations between the X-ray luminosity and the spin period derivative strongly suggest the presence
of an accretion disc around GRO~J1008$-$57 during the X-ray outbursts \citep[see, e.g., ][]{Wang21}.
The distance of the system is either $9.7\pm0.8$~kpc or $5.8\pm0.5$~kpc \citep{Riquelme12}.

\citet{Xing19} analysed nine years of \emph{Fermi}/LAT data and used the 4-yr LAT catalogue.
They reported the discovery of a transient $\gamma-$ray emission spatially coincident with GRO~J1008$-$57.
The $\gamma-$ray emission is only observed at the orbital phase 0.8$-$0.9 (where phase 0 is the periastron),
and the detection is due mostly to three events, one of which, in 2012-2013, after a giant X-ray outburst.
The $\gamma-$ray luminosity was $L_\gamma=3.6\times 10^{34}$~erg~s$^{-1}$ in 0.5$-$39~GeV
(or $L_\gamma=6.7\times 10^{34}$~erg~s$^{-1}$ in 0.1$-$100~GeV).
The X-ray luminosity during the \emph{Fermi}/LAT detections was likely around $L_{\rm x}\approx10^{35}$~erg~s$^{-1}$ \citep{Xing19}.
\citet{Harvey22} analysed 12 years of \emph{Fermi}/LAT data. They detected GRO~J1008$-$57
($z=4.9\sigma$) with a flux of $\sim 3.24\times 10^{-6}$~MeV~cm$^{-2}$~s$^{-1}$ (0.1$-$500~GeV).
The lightcurve folded with the orbital period showed excesses at three orbital phases, with two bins
consistent with the results of \citet{Xing19}.
\citet{Harvey22} do not claim the detection of this source, being all the flux measurements below 5$\sigma$,
but pointed out that it is unlikely that the points with higher flux cluster by chance in phase space.

\begin{figure}
\begin{center}
  \includegraphics[width=8.5cm]{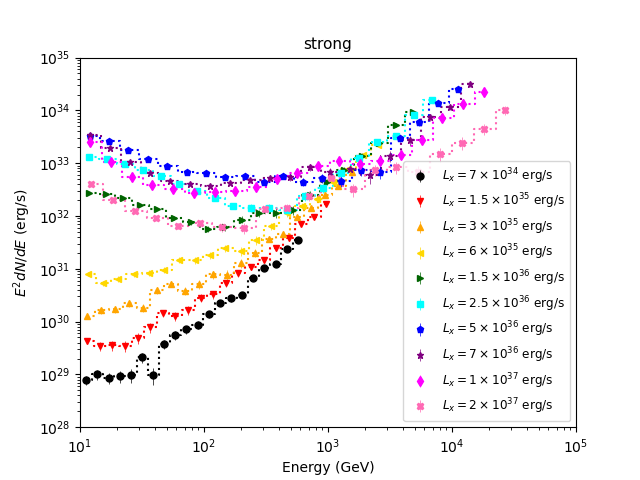}
  \includegraphics[width=8.5cm]{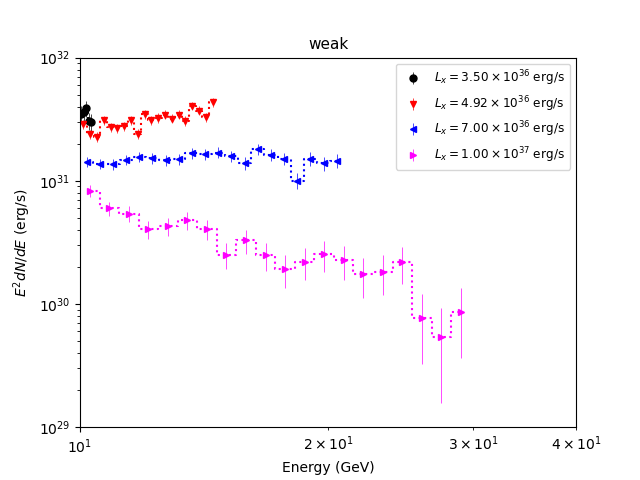}
\end{center}
\caption{$\gamma$-ray spectra of GRO~J1008$-$57 for the photons which reached the observer, assuming strong (left panel) and weak (right panel) shielding.}
\label{fig:gro spectra}
\end{figure}

Similarly to the case of A0535+26, and with the aim of having a spectra grid that can be used for comparisons with future observations,
we simulated the $\gamma-$ray spectra of GRO~J1008$-$57 for different 
input values of $L_{\rm x}$ (that varies from $L_{\rm x}=10^{34}$~erg~s$^{-1}$ to $L_{\rm x}= 2\times 10^{37}$~erg~s$^{-1}$),
assuming strong and weak shielding of the gap. The spectra are shown in Fig. \ref{fig:gro spectra}.
The high $\gamma-$ray luminosity reported in \citet{Xing19} and \citet{Harvey22}, $L_\gamma\approx 10^{34}$~erg~s$^{-1}$,
likely corresponding to an X-ray luminosity state of $L_{\rm x}\approx 10^{34-35}$~erg~s$^{-1}$, could be roughly explained
assuming a strong shielded gap and the 3$\sigma$ upper limit of $\approx 0.5$\% for the beaming factor at $L_{\rm x}\approx 10^{35}$~erg~s$^{-1}$.
For low $L_{\rm x}$, we indeed obtain an upper-limit for $f_{\rm b}$ of the order of $\approx 0.5$\%.

\section{Conclusions and future work}

We explored the properties of the $\gamma-$ray emission produced by a pulsar in an XRB, fed by an accretion disc.
Our calculations stem from the model presented by \citetalias{Cheng89} and they take into account 
several physical processes for the development of photon-electron cascades, inside and outside the accretion disc.
Compared to previous works based on \citetalias{Cheng89}, for the cascades we considered more physical processes, which involve
interactions with nuclei, soft photons from the accretion disc, and the magnetic field.
Compared to the work by \citet{Orellana07}, we have also considered synchrotron, curvature, and magnetic pair production.
For the first time, we thoroughly explored the case of ``weak shielding''
and we considered a more complex configuration for the magnetic fields inside the accretion disc
that takes into account the poloidal component.
We also presented a wide grid of solutions, which can be used for comparison purposes on many XRBs
having a broad spectrum of properties.
The grids of solutions are based on different input parameter values of the X-ray luminosity ($L_{\rm x}$),
magnetic field strength ($B$), and for different properties of the region where acceleration occurs,
namely ``strong'' and ``weak'' shielding in the gap.
We explored the parameter space $8\times 10^{34} \leq L_{\rm x} \leq 5\times 10^{37}$~erg~s$^{-1}$,
$10^{11} \leq B \leq 8 \times 10^{12}$~G.
We found that the $\gamma-$ray luminosity spans more than five orders of magnitude, with a maximum of $\sim10^{35}$~erg~s$^{-1}$.
The $\gamma-$ray spectra show a large variety of shapes: some have most of the emission below $\sim100$~GeV (weak shielding),
others are harder, with emission up to 10$-$100~TeV (strong shielding).
The beaming factor varies from $\lesssim 0.01$ to $\sim 0.35$.

We compared our results with \emph{Fermi}/LAT detections and VERITAS upper-limits
of two HMXBs: A0535+26 and GRO~J1008$-$57. More consequential comparisons will be possible when
more sensitive instruments will be operational in the coming years,
such as, for example, CTAO, ASTRI Mini-Array,
the Southern Wide-field Gamma-ray Observatory (SWGO),
  and the High Energy cosmic Radiation Detection facility (HERD)
\citep{Zhang14b, Huentemeyer19}.

The results shown in this work are based on a number of simplifications, which will be addressed in future works.
We list the ones we consider most significant.
The model by \citetalias{Cheng89} and our calculations are based on the assumption that the rotational and dipole magnetic axes
are aligned. This assumption greatly simplifies the calculations, as demonstrated by other numerous works on pulsars which adopt it.
Nevertheless, it is inevitable that our solutions may diverge from those of the more realistic case of an oblique rotator.
A similar reasoning applies to the curvature radiation produced within the disc, which we have neglected and could be included in a future work.
The interaction effects of the radiation field from the companion star could also be included.
\citet{Bednarek93, Bednarek97, Bednarek00, Sierpowska05} showed that if it is taken into account,
the observed $\gamma-$ray emission can show an orbital modulation.
Also the beaming factor can be affected by the assumptions of aligned rotators and the geometry
of the magnetic field inside the accretion disc. Therefore, the values reported in this work should be considered within the bounds of the assumptions under which they were derived.
Finally, we have considered here only a model, based on the that presented by \citetalias{Cheng89}.
In Sect. \ref{Sect. Models} we briefly listed
many other models on which work similar to this could be done,
and whose importance will become more and more urgent in view of future observational advances
in the $\gamma-$ray energy band.

\section*{Acknowledgements}

We thank the anonymous referee for constructive comments that helped to improve the paper.
The authors acknowledge support by the High Performance and Cloud Computing Group
at the Zentrum f\"ur Datenverarbeitung of the University of T\"ubingen,
the state of Baden-W\"urttemberg through bwHPC and the German Research Foundation (DFG)
through grant no INST 37/935-1 FUGG.
We acknowledge financial contribution from the agreement ASI-INAF I/037/12/0 and ASI-INAF n. 2017-14-H.0.
L.D. thanks Andrea Tramacere, Denys Malyshev, and Samuele Chimento for the useful advice that helped improve some aspects of this work.

\section*{Data Availability}

Data of specific test simulations can be obtained upon reasonable
request from the corresponding author.



\bibliographystyle{mnras}
\bibliography{ldvhe} 




\appendix

\section{General outline of the procedure to calculate the observable $\gamma-$ray emission}
\label{sect. appendix computational procedure}

Figure \ref{fig. flow-chart-A} shows a flow-chart to describe the main aspects of the procedure used to calculate the 
  $\gamma-$ray emission from an XRB. It can be divided in five steps:
  \begin{enumerate}
  \item input parameters;
  \item calculation of the proton energies and production of $\gamma-$ray primary photons inside the accretion disc;
  \item calculation of the cascades produced inside of the accretion disc;
  \item calculation of the cascades produced outside the accretion disc;
    \item Output of the list of photons arriving at the observer, the expected observed flux and spectrum above 10~GeV.
  \end{enumerate}
  Other details are given in Sect. \ref{sect. cascades}.

\begin{figure*}
  \begin{center}
      \includegraphics[width=15cm]{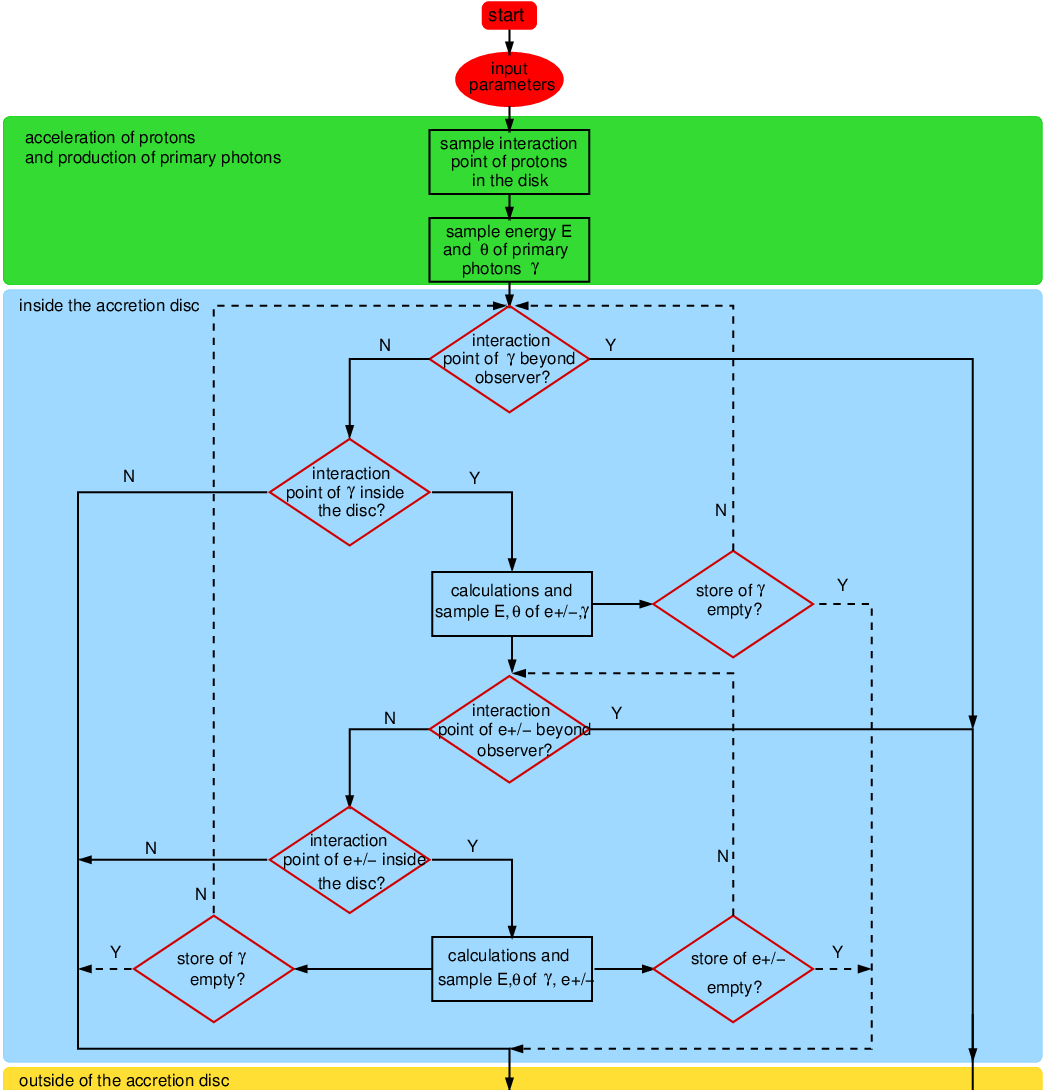}
\end{center}
\caption{Schematic representation illustrating the computational procedure for calculating
  the expected observable $\gamma-ray$ emission (for the second part of the flow-chart, see next figure).
  Some lines are dotted for ease of understanding when they cross other lines.}
\label{fig. flow-chart-A}
\end{figure*}

\begin{figure*}
  \begin{center}
      \includegraphics[width=15cm]{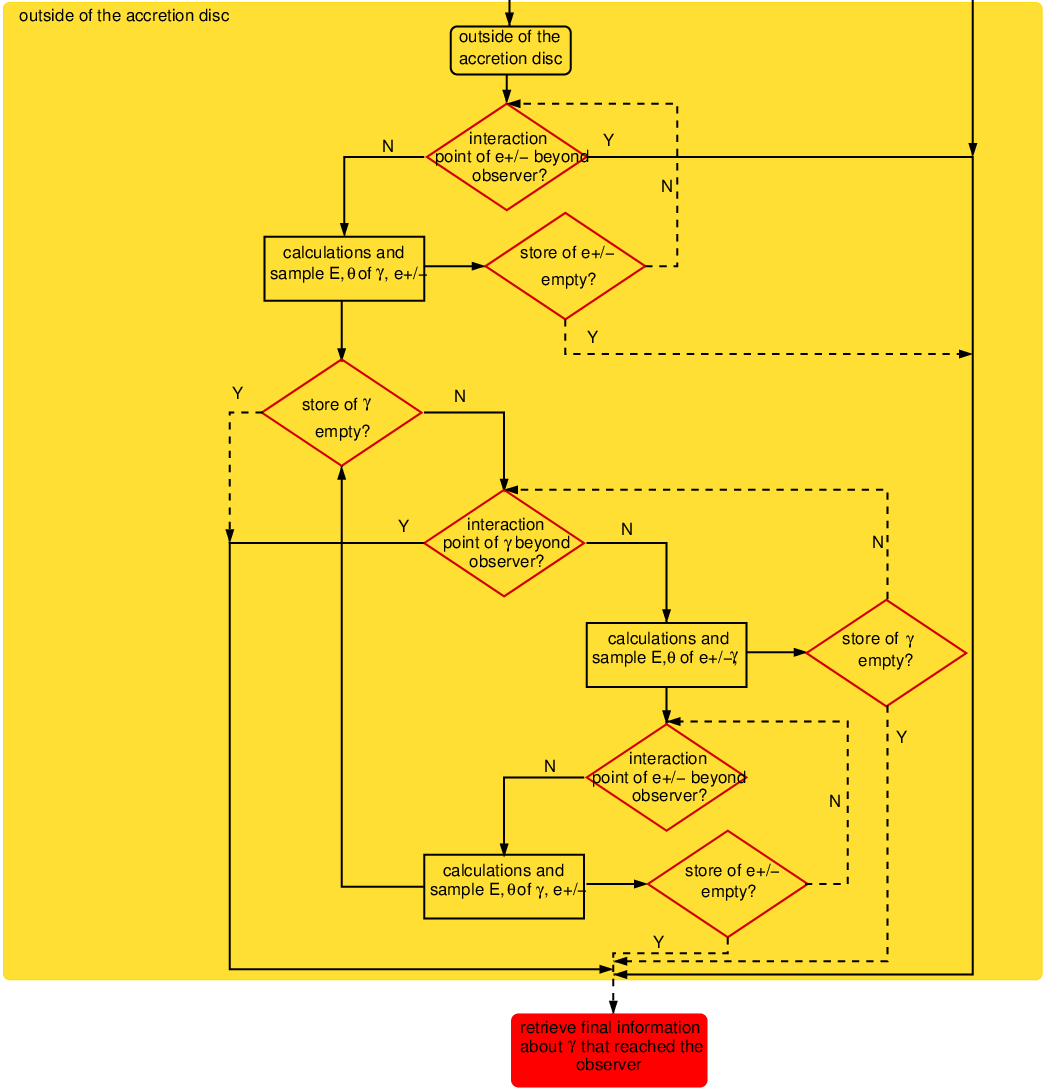}
\end{center}
\contcaption{Second part of the schematic representation illustrating the computational procedure for calculating
  the expected observable $\gamma-$ray emission.}
\label{fig. flow-chart-B}
\end{figure*}

\clearpage
\newpage

\section{Optical depth for $\gamma-$ray photons in the photon field of a massive star}
\label{appendix:tau}

\begin{figure}
\begin{center}
  \includegraphics[width=9cm]{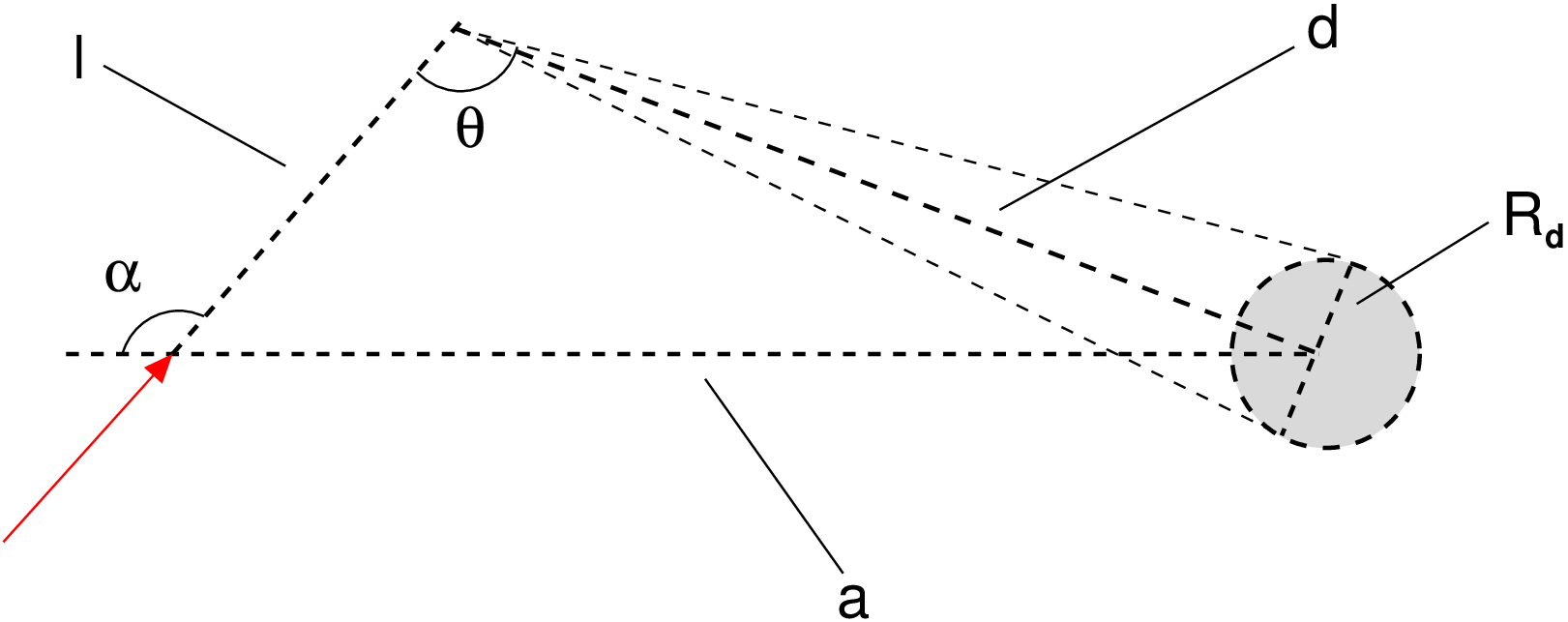}
\end{center}
\caption{Schematic view of a binary system composed by a NS, where a $\gamma-$ray photon (red arrow) is injected with an angle $\alpha$, and a donor star
  with radius $R_{\rm d}$. The stars are separated by a distance $a$.}
\label{fig:tau_vs_alpha_xfig}
\end{figure}

Following an approach similar to that described in \citet{Bednarek97}, we assume that a $\gamma-$ray photon with
energy $E_\gamma$ is injected with an angle $\alpha$ at a distance $a$ from the center of the companion star (Fig. \ref{fig:tau_vs_alpha_xfig}).
Following \citet{Gould67} (their eq. 7), the optical depth for the propagation of the $\gamma-$ray photon in the radiation field
of the donor star is given by:
\begin{equation} \label{eq.tau}
\tau = \int_{R_0}^{D} \int_{\theta_{\rm min}}^{\theta_{\rm max}} \int_{\varepsilon(s>1)}^{\infty} \frac{1}{2}\sigma_{\gamma\gamma}(\beta)n(\varepsilon, d(l))(1-\cos\theta)\sin\theta d\varepsilon d\theta dl \mbox{ ,}
\end{equation}
where $D$ is the distance from the observer, $R_0$ is the inner radius of the accretion disc around the NS,
$\theta$ is the angle of interaction between the $\gamma-$ray photon and the soft photons, $n(\epsilon, d)$ is the
blackbody photon number density at a distance $d$ from the donor star,
$\sigma_{\gamma\gamma}(\beta)$ is the cross section for $e^\pm$ pair production in two photon collision
\citep[][ and references therein]{Gould67}. $\beta$ is given by:
\begin{eqnarray}
  \beta & = & \sqrt{1-\frac{1}{s}} \nonumber \\
  s     & = & \frac{\varepsilon E_\gamma}{2m_e^2c^4} (1 - \cos \theta) \nonumber \mbox{ ,}
\end{eqnarray}
Obviously, for pair production to occur, $s>1$.
The angle of interaction $\theta$ varies from $\theta_{\rm min}$ and $\theta_{\rm max}$, due to the size of the donor star.
These angles are given by:
\begin{eqnarray}
  \theta & = & \arctan \left( \frac{a - l \cos (\pi - \alpha)}{l \sin (\pi - \alpha)} \right) + \alpha - \frac{\pi}{2} \nonumber \\
  \theta_{\rm min} & = & \theta - \arctan \left( \frac{R_{\rm d}}{d} \right) \nonumber  \\
  \theta_{\rm max} & = & \theta + \arctan \left( \frac{R_{\rm d}}{d} \right) \nonumber 
\end{eqnarray}
Equation (\ref{eq.tau}) provides only a rough estimate of the optical depth. For example,
we neglected the spherical shape of the donor star, being, for the cases of our interest, $a\gg R_{\rm d}$.
A detailed study (whose solutions are valid for binary systems with narrower orbits than those considered in this work) can be found in \citet{Bednarek97}.

\section{Contribution of different interaction processes to the $\gamma-$ray spectrum: weak shielding case}
\label{sect. appendix}

\begin{figure*}
\begin{center}
  \begin{tabular}{lccc}
                                                                     & Case A & Case B & Case C \\
\rotatebox{90}{\hspace{0.5cm} $L_{\rm x}=3\times10^{36}$~erg~s$^{-1}$} & \includegraphics[width=5.5cm]{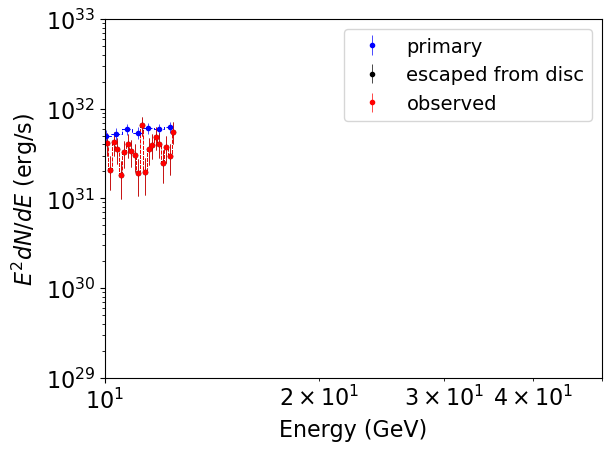} & \includegraphics[width=5.5cm]{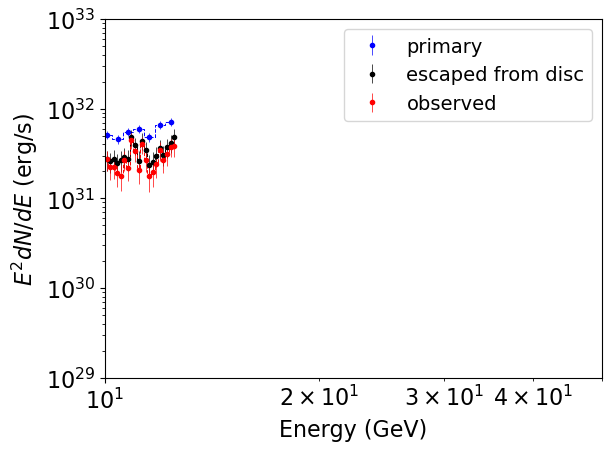} & \includegraphics[width=5.5cm]{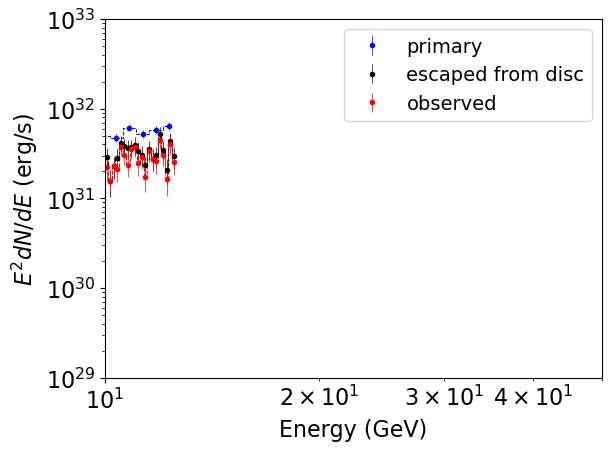}\\
\rotatebox{90}{\hspace{0.5cm} $L_{\rm x}=5\times10^{36}$~erg~s$^{-1}$} & \includegraphics[width=5.5cm]{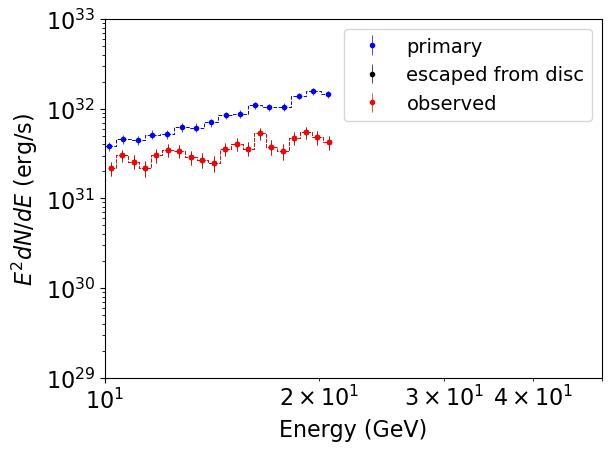} & \includegraphics[width=5.5cm]{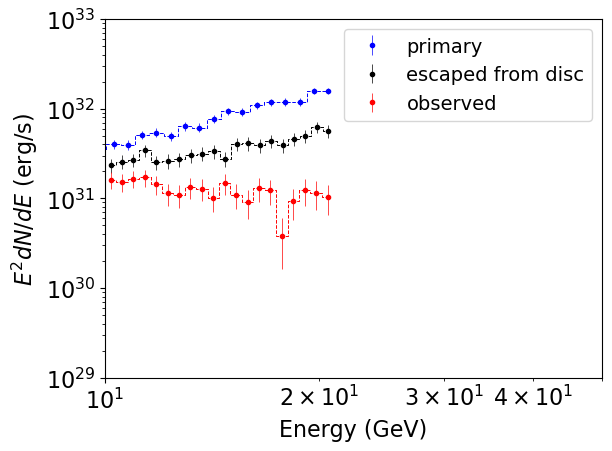} & \includegraphics[width=5.5cm]{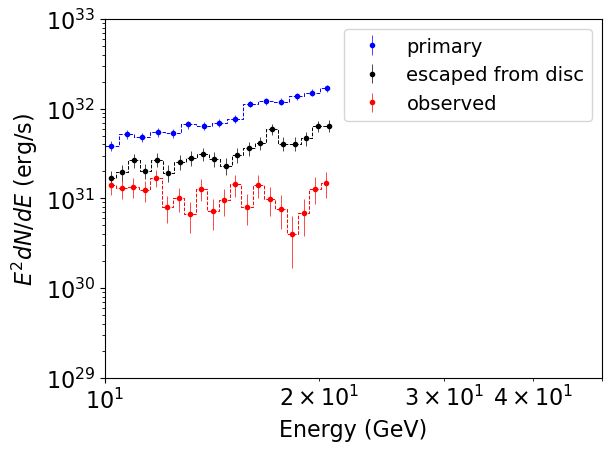}\\
\rotatebox{90}{\hspace{0.5cm} $L_{\rm x}=7\times 10^{36}$~erg~s$^{-1}$} & \includegraphics[width=5.5cm]{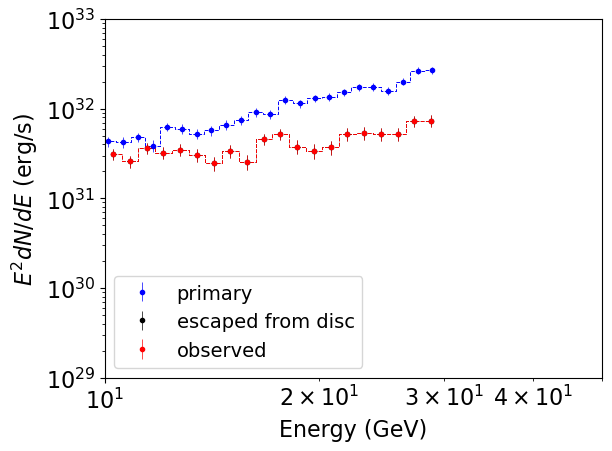} & \includegraphics[width=5.5cm]{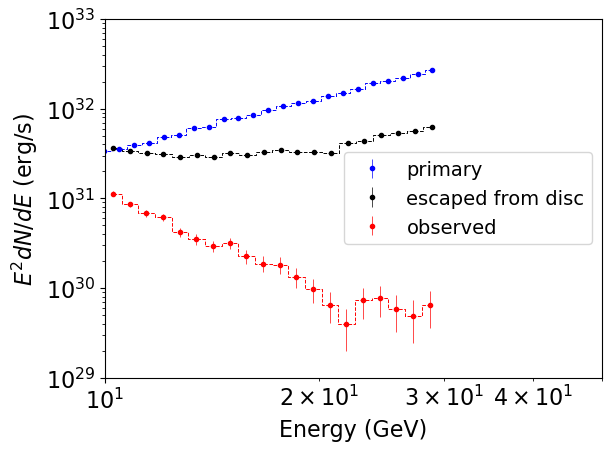} & \includegraphics[width=5.5cm]{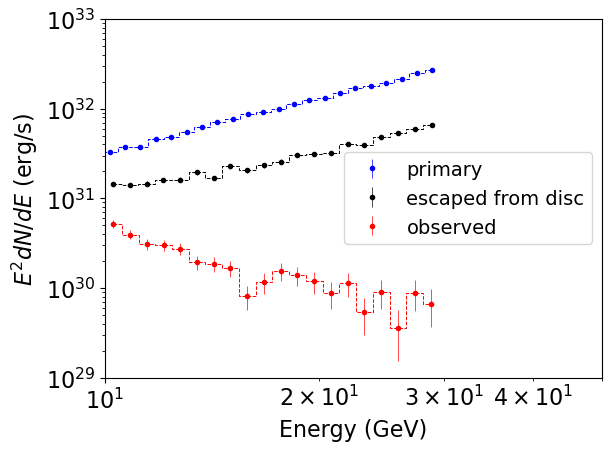}\\
\rotatebox{90}{\hspace{0.5cm} $L_{\rm x}=1\times 10^{37}$~erg~s$^{-1}$} & \includegraphics[width=5.5cm]{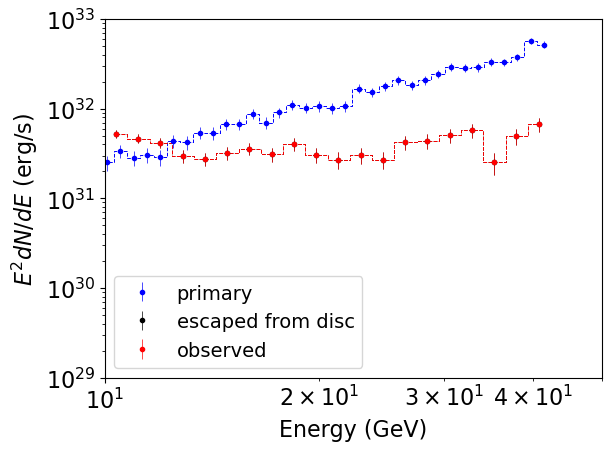} & \includegraphics[width=5.5cm]{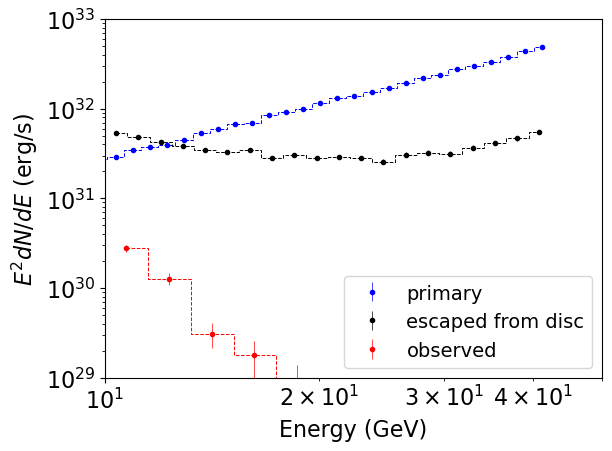} & \includegraphics[width=5.5cm]{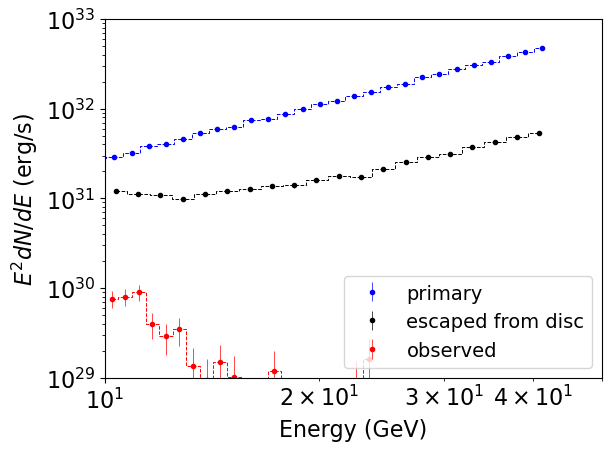}\\
\end{tabular}
\end{center}
\caption{$\gamma$-ray spectra for the primary photons,
  photons which escaped from the disc, and photons which reached the observer,
  for different X-ray luminosities, and for different radiation and pair processes activated, as described in Table \ref{tab:a0535_expl_we}.
  We assumed $B=4\times 10^{12}$~G (i.e. a framework compatible with the Be/XRB A0535+26) and a weak shielded gap.}
\label{fig:a0535_expl_we}
\end{figure*}

\begin{figure*}
\begin{center}
  \begin{tabular}{lccc}
                                                                     & Case A & Case B & Case C \\
\rotatebox{90}{\hspace{1.0cm} $L_{\rm x}=3\times10^{36}$~erg~s$^{-1}$} & \includegraphics[width=4.5cm]{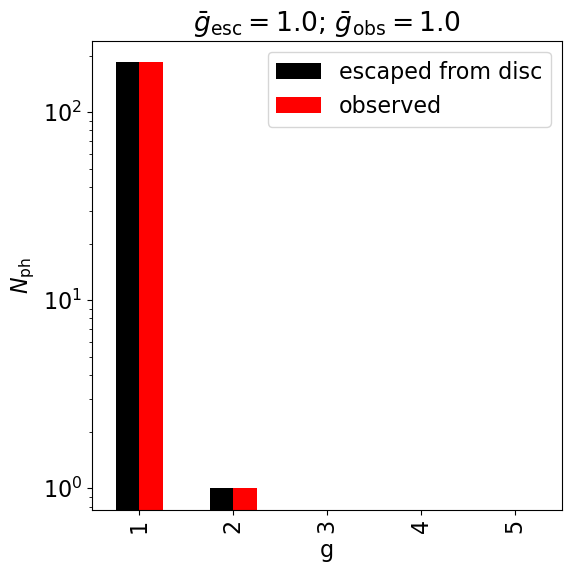} & \includegraphics[width=4.5cm]{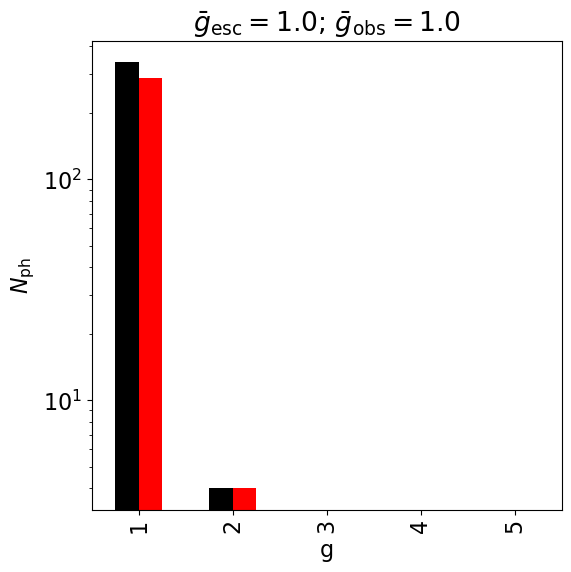} & \includegraphics[width=4.5cm]{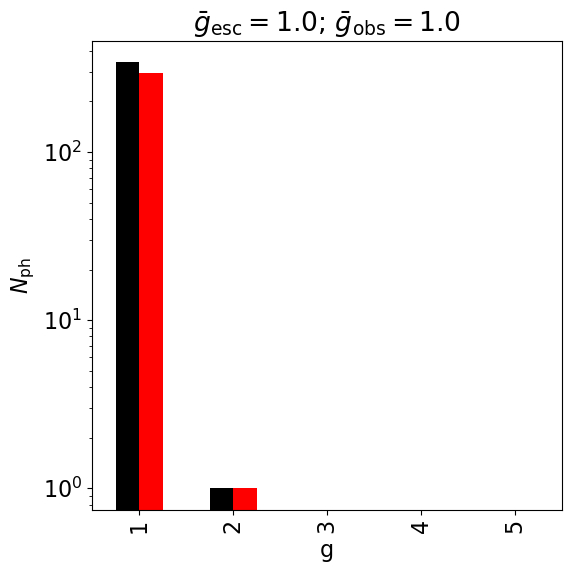}\\
\rotatebox{90}{\hspace{1.0cm} $L_{\rm x}=5\times10^{36}$~erg~s$^{-1}$} & \includegraphics[width=4.5cm]{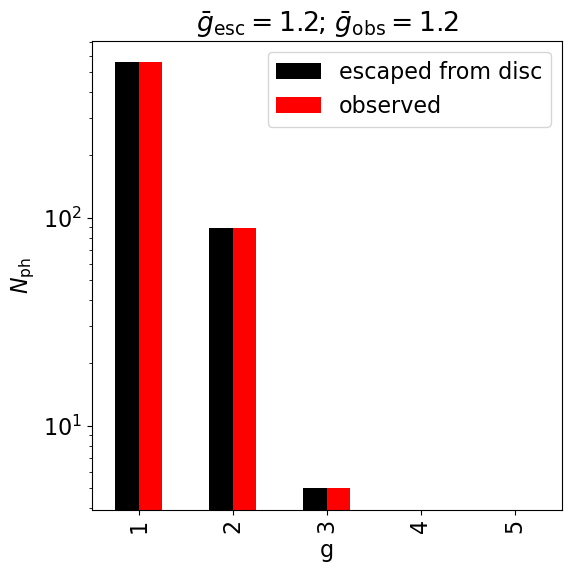} & \includegraphics[width=4.5cm]{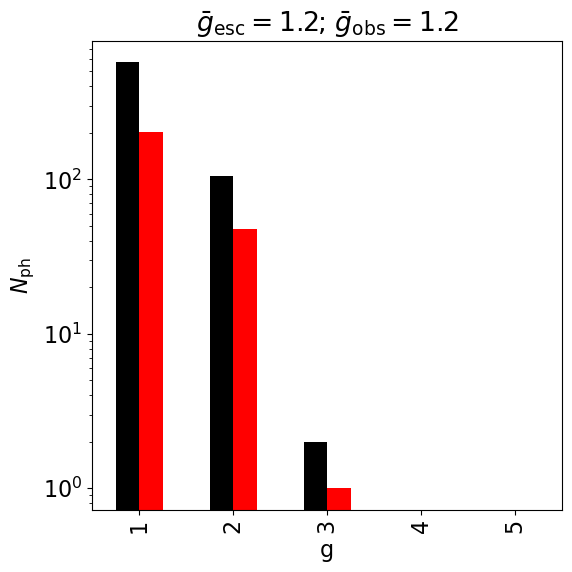} & \includegraphics[width=4.5cm]{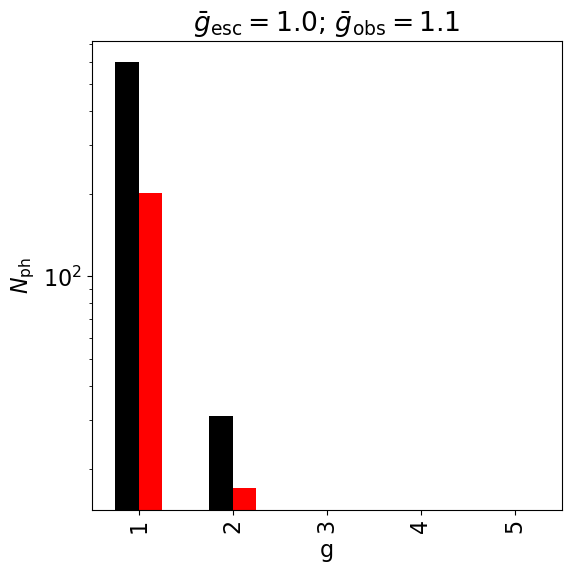}\\
\rotatebox{90}{\hspace{1.0cm} $L_{\rm x}=7\times 10^{36}$~erg~s$^{-1}$} & \includegraphics[width=4.5cm]{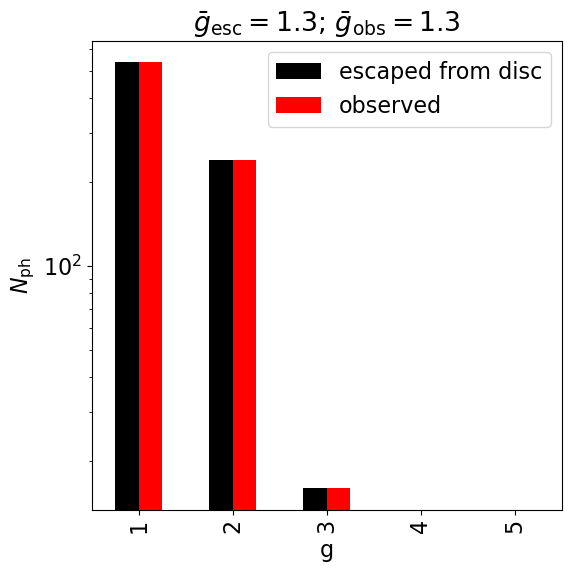} & \includegraphics[width=4.5cm]{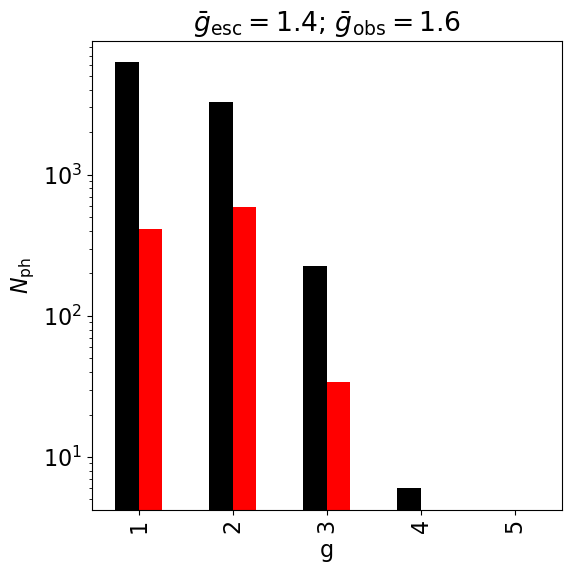} & \includegraphics[width=4.5cm]{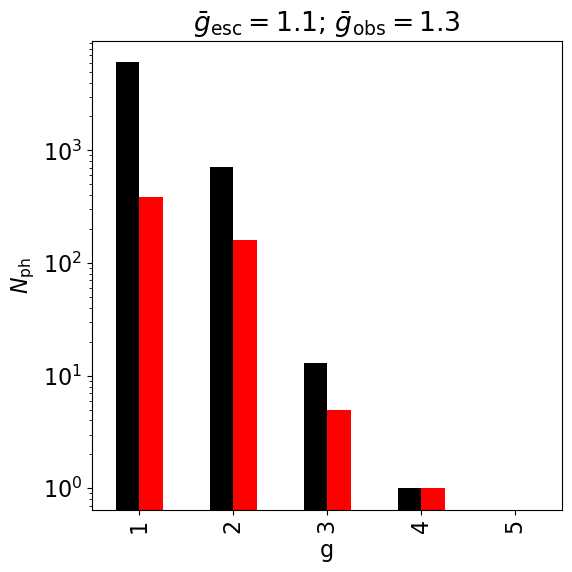}\\
\rotatebox{90}{\hspace{1.0cm} $L_{\rm x}=1\times 10^{37}$~erg~s$^{-1}$} & \includegraphics[width=4.5cm]{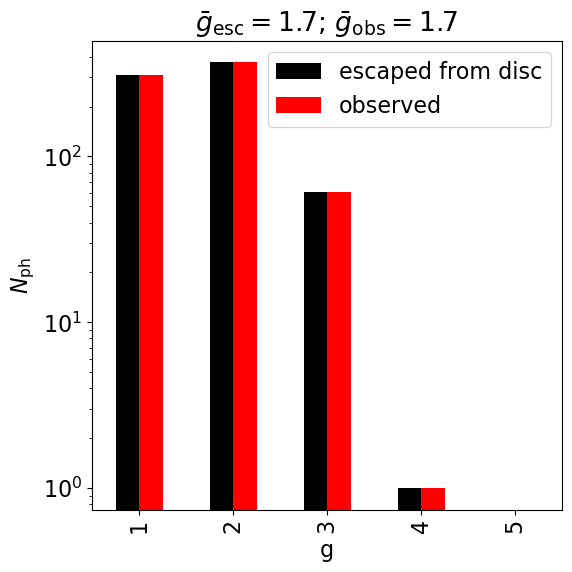} & \includegraphics[width=4.5cm]{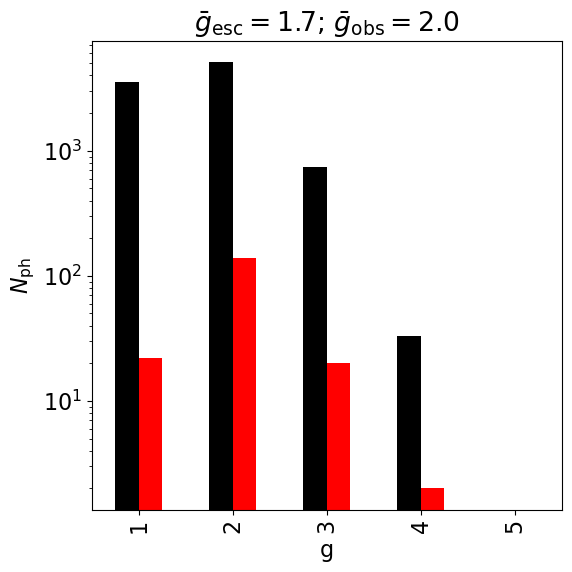} & \includegraphics[width=4.5cm]{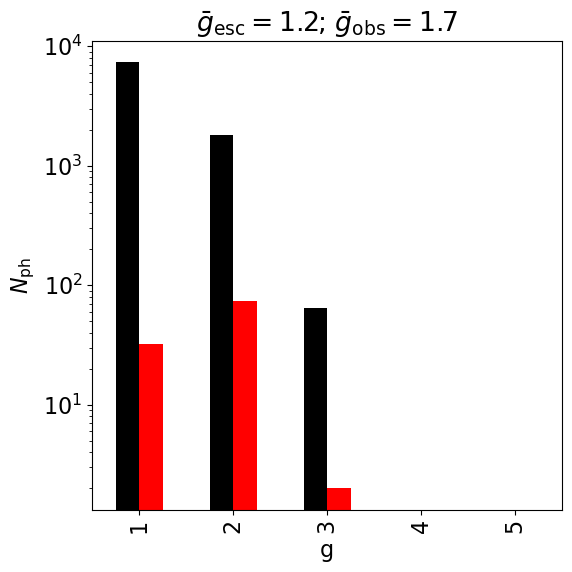}\\
\end{tabular}
\end{center}
\caption{Histograms of the generations of photons that escape from the disc (black histogram) and reach the observer (red histogram), for different X-ray luminosities
  and radiation and pair processes activated as described in Table \ref{tab:a0535_expl}.
  We assumed $B=4\times 10^{12}$~G (i.e. a framework compatible with the Be/XRB A0535+26) and a weak shielded gap.
  On top of each panel is shown the average value of the generation number for photons that escaped the disc ($\bar{g}_{\rm esc}$)
  and those that reached the observer ($\bar{g}_{\rm obs}$).}
\label{fig:a0535_expl_gen_we}
\end{figure*}

Table \ref{tab:a0535_expl_we} and Fig. \ref{tab:a0535_expl_we} show
the observable $\gamma-$ray luminosities and spectra (above 10~GeV)
calculated assuming a weakly shielded gap, four different X-ray luminosities, and different types of interactions processes activated.
In particular, Fig. \ref{tab:a0535_expl_we} shows
the $\gamma-$ray spectra (above 10~GeV) from photons that escaped from the accretion disc, those observed
(i.e. photons that escaped from the disc and survived along their journey between the source and the observer),
and the spectrum of the primary photons (those created by the $\pi^0$ decay inside the disc).
Figure \ref{fig:a0535_expl_gen_we} shows, for each of the simulations displayed in Fig. \ref{fig:a0535_expl_we},
how the cascade of $\gamma-$ray photons develops inside and outside the disc.
Similar results for the strong shielding case are described in Sect. \ref{Sect. contrib. interactions}.

\begin{table*}
\begin{center}
\caption{$\gamma$-ray luminosities ($\geq 10$~GeV), for the primary photons ($L_{\rm pri}$),
  photons which escaped from the disc ($L_{\rm esc}$), and photons which reached the observer ($L_{\rm obs}$),
  assuming weak shielding, for different X-ray luminosities, and for different radiation and pair processes activated.
  We assumed $B=4\times 10^{12}$~G (i.e. a framework compatible with the Be/XRB A0535+26).}
\label{tab:a0535_expl_we}
\begin{tabular}{lcccc} 
\hline
$L_{\rm x}$       & $L_\gamma$   &                Case A$^a$    &       Case B$^b$               &        Case C$^c$              \\
\hline
                &             &          nuclei             &           nuclei               &              nuclei            \\
                &             &                             &         + photons               &         + photons               \\
                &             &                              &                                 &          + $\vec{B}$            \\
\hline
erg~s$^{-1}$     &             &        erg~s$^{-1}$          &          erg~s$^{-1}$           &        erg~s$^{-1}$             \\
\hline
$3\times10^{36}$  & $L_{\rm pri}$ & $1.37 \pm 0.20 \times 10^{31}$ &  $1.36 \pm 0.14 \times 10^{31}$   &   $1.21 \pm 0.16 \times 10^{31}$  \\
                 & $L_{\rm esc}$ & $8.08 \pm 2.62 \times 10^{30}$ &  $7.52 \pm 1.80 \times 10^{30}$   &   $7.45 \pm 1.78 \times 10^{30}$  \\
                 & $L_{\rm obs}$ & $8.08 \pm 2.62 \times 10^{30}$ &  $6.35 \pm 1.65 \times 10^{30}$   &   $6.43 \pm 1.66 \times 10^{30}$  \\
\hline
$5\times10^{36}$  & $L_{\rm pri}$ & $6.44\pm 0.64\times 10^{31}$ &  $6.51 \pm 0.64\times 10^{31}$   &   $6.68\pm 0.62\times 10^{31}$  \\
                 & $L_{\rm esc}$ & $2.62\pm 0.46\times 10^{31}$ &  $2.71 \pm 0.46 \times 10^{31}$  &   $2.57 \pm 0.45 \times 10^{31}$  \\
                 & $L_{\rm obs}$ & $2.62\pm 0.46\times 10^{31}$ &  $9.07 \pm 2.61 \times 10^{30}$  &   $7.96 \pm 2.44 \times 10^{30}$  \\
\hline
$7\times10^{36}$  & $L_{\rm pri}$ & $1.25\pm 0.13\times 10^{32}$ &  $1.26\pm 0.14\times 10^{32}$   &   $1.27\pm 0.13\times 10^{32}$  \\
                 & $L_{\rm esc}$ & $4.54\pm 0.72\times 10^{31}$ &  $4.19\pm 0.69\times 10^{31}$   &   $3.19 \pm 0.61 \times 10^{31}$  \\
                 & $L_{\rm obs}$ & $4.54\pm 0.72\times 10^{31}$ &  $3.00 \pm 1.25 \times 10^{30}$ &   $1.88 \pm 1.03 \times 10^{30}$  \\
\hline
$10^{37}$         & $L_{\rm pri}$ & $2.39\pm 0.26\times 10^{32}$ &  $2.36\pm 0.44\times 10^{32}$   &   $2.36\pm 0.44\times 10^{32}$  \\
                 & $L_{\rm esc}$ & $5.67\pm 0.98\times 10^{31}$ &  $5.26\pm 0.26\times 10^{31}$   &   $3.22 \pm 0.15 \times 10^{31}$  \\
                 & $L_{\rm obs}$ & $5.67\pm 0.98\times 10^{31}$ &  $6.43 \pm 1.00\times 10^{29}$  &   $1.89 \pm 0.63 \times 10^{29}$  \\
\hline
\end{tabular}
\end{center}
\begin{flushleft}
  $^a$ Case A: interaction with nuclei: pair production in the Coulomb field of nucleus and bremsstrahlung.\\
  $^b$ Case B: interaction with nuclei and photons from the accretion disc: pair production in the Coulomb field of nucleus, pair production in photon-photon collisions, bremsstrahlung, inverse Compton with soft photons from the accretion disc.\\
  $^c$ Case C: interaction with nuclei and photons from the accretion disc, and with the magnetic field: pair production in the Coulomb field of nucleus, pair production in photon-photon collisions, magnetic pair production, bremsstrahlung, inverse Compton with soft photons from the accretion disc, synchrotron, and curvature.
\end{flushleft}
\end{table*}

\bsp	
\label{lastpage}
\end{document}